\documentclass[useAMS]{mn2e}

\usepackage{graphicx}

\title[Spectroscopic study of red giants in the {\it Kepler} field]
{Spectroscopic study of red giants in the {\it Kepler} field with  
asteroseismologically established evolutionary status and stellar parameters 
}

\author[Y. Takeda and A. Tajitsu]
{Y. Takeda$^{1}$\thanks{E-mail:
takeda.yoichi@nao.ac.jp}\footnotemark[0]\thanks{
Based on data collected at Subaru Telescope, which is operated by the 
National Astronomical Observatory of Japan.} and
A. Tajitsu$^{2}$\\
$^{1}$National Astronomical Observatory of Japan, 
2-21-1 Osawa, Mitaka, Tokyo 181-8588, Japan\\
$^{2}$Subaru Telescope, 650 N. A'ohoku Place, Hilo, HI 96720, U.S.A.\\
}
\begin{document}

\date{Accepted 2015 March 25. Received 2015 March 5; in original form 2015 February 3}


\maketitle

\label{firstpage}

\begin{abstract}
Thanks to the recent very high-precision photometry of red giants from satellites
such as $Kepler$, precise mass and radius values as well as accurate information 
of evolutionary stages are already established by asteroseismic approach for a 
large number of G--K giants.
Based on the high-dispersion spectra of selected such 55 red giants in the $Kepler$ 
field with precisely known seismic parameters (among which parallaxes are available 
for 9 stars), we checked the accuracy of the determination method of stellar parameters 
previously applied to many red giants by Takeda et al. (2008, PASJ, 60, 781), 
since it may be possible to discriminate their complex evolutionary status 
by using the surface gravity vs. mass diagram. 
We confirmed that our spectroscopic gravity and the seismic 
gravity satisfactorily agree with each other (to within
$\simeq 0.1$~dex) without any systematic difference. 
However, the mass values of He-burning red clump giants derived from stellar 
evolutionary tracks ($\sim$~2--3~M$_{\odot}$) were found to be markedly larger  
by $\sim$~50\% compared to the seismic values ($\sim$~1--2~M$_{\odot}$) 
though such discrepancy is not seen for normal giants in the H-burning phase, 
which reflects the difficulty of mass determination from intricately overlapping tracks
on the luminosity vs. effective temperature diagram. 
This consequence implies that the mass results of many red giants in the clump region 
determined by Takeda et al. (2008) are likely to be significantly overestimated. 
We also compare our spectroscopically established parameters with recent literature 
values, and further discuss the prospect of distinguishing the evolutionary
status of red giants based on the conventional (i.e., non-seismic) approach.
\end{abstract}


\section{Introduction}

Red giant stars are low-to-intermediate mass ($\sim$~1--5~M$_{\odot}$) stars
which have already evolved off the main sequence with lowered temperature
as well as inflated radius, currently situating in the high luminosity 
($L \sim$10--1000~L$_{\odot}$) and late spectral type 
($T_{\rm eff} \sim$~4000--5500~K) region in the HR diagram.
Since they are intrinsically bright and numerous in number while covering 
a wide range of age, their astrophysical importance is widely known
(e.g., as probes of galactic chemical evolution).

\setcounter{figure}{0}
\begin{figure}
\begin{minipage}{80mm}
\includegraphics[width=8.0cm]{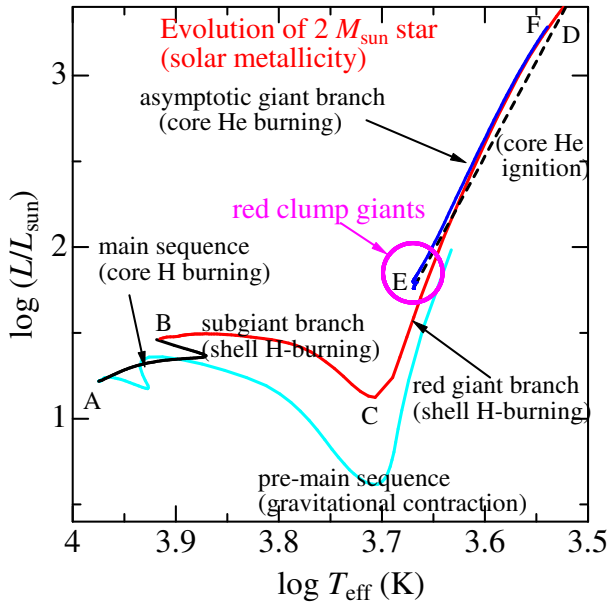}
\caption{Typical evolutionary track of red giants on the $\log L$ vs. 
$\log T_{\rm eff}$ diagram. Shown here is the case of
a 2~M$_{\odot}$ star of solar metallicity calculated by Lagarde et al. (2012).}
\label{fig1}
\end{minipage}
\end{figure}

One notorious problem regarding red giants is that an inevitable ambiguity 
is involved in understanding their evolutionary status. As shown in Fig. 1,
normal giants of the red giant branch (RGB) in the shell H-burning phase 
(C$\rightarrow$D) and the He-core burning giants of the asymptotic giant 
branch (AGB: E$\rightarrow$F and further) follow almost same track in the HR diagram, 
which makes understanding the evolutionary status of a given star very difficult 
based on the position on this diagram alone. 
Especially important are the red-clump (RC) giants (around E), which have 
just started stable core He burning after the He-ignition (D$\rightarrow$E) 
and are bound to gradually ascend the AGB. They tend to cluster at a fixed 
luminosity because of their comparatively slow evolution, and thus making up 
a considerably large fraction of red giants.
But it is very hard to tell only from the HR diagram whether a given star in this 
red clump region is a real RC star or a normal RGB star just wandering into this area.

Fortunately, recent progress in asteroseismology has shed light on this 
stagnant situation. The extremely high-precision photometry accomplished by 
satellite observations (such as $CoRoT$ or $Kepler$) has made it possible to detect 
very subtle photometric variability of red giants pulsating in the mixed p- and g-mode.
It then revealed that normal giants and red-clump giants were clearly
distinguished (by using the $\Delta \nu$ vs. $\Delta \Pi_{1}$ diagram) 
and mass as well as radius could be precisely determined (from $\nu_{\rm max}$ 
and $\Delta \nu$ with the help of the scaling relations) by analyzing 
the power spectrum of their oscillation.
In this way, Mosser et al. (2012) successfully classified many G--K giants 
in the {\it Kepler} field into three categories of (a) normal red giants
(RG: before He ignition), (b) 1st clump giants (RC1: ordinary RC stars of 
comparatively lower mass after degenerate He-ignition), and (c) 2nd clump giants 
(RC2: RC stars of comparatively higher mass after non-degenerate He-ignition; 
cf. Girardi 1999), and published their asteroseismic radius ($R_{\rm seis}$) 
and mass ($M_{\rm seis}$) values. 

Yet, while such an achievement of distinguishing RG/RC1/RC2 by way of asteroseismology 
is surely a great benefit for the astronomical community, such a special technique 
of very high-precision photometry is not always applicable. Is it possible to 
find a way based on the conventional method to discriminate the evolutionary status 
of red giants in general?

Takeda, Sato \& Murata (2008, hereinafter referred to as Paper I) conducted an extensive 
stellar parameter study for apparently bright ($V <6.5$) 322 field giants, 
for which atmospheric parameters [the effective temperature ($T_{\rm eff}$), 
surface gravity ($\log g$), microturbulence ($v_{\rm t}$), and metallicity ([Fe/H])] 
were spectroscopically determined from Fe~{\sc i} and Fe~{\sc ii} lines. 
The stellar age ($age$) and mass ($M$) were derived by comparing their positions 
on the HR diagram ($\log L$ vs. $\log T_{\rm eff}$) with theoretical evolutionary tracks,
where the luminosity ($L$) was evaluated with the help of Hipparcos parallaxes (ESA 1997).

\setcounter{figure}{1}
\begin{figure*}
\begin{minipage}{130mm}
\includegraphics[width=13.0cm]{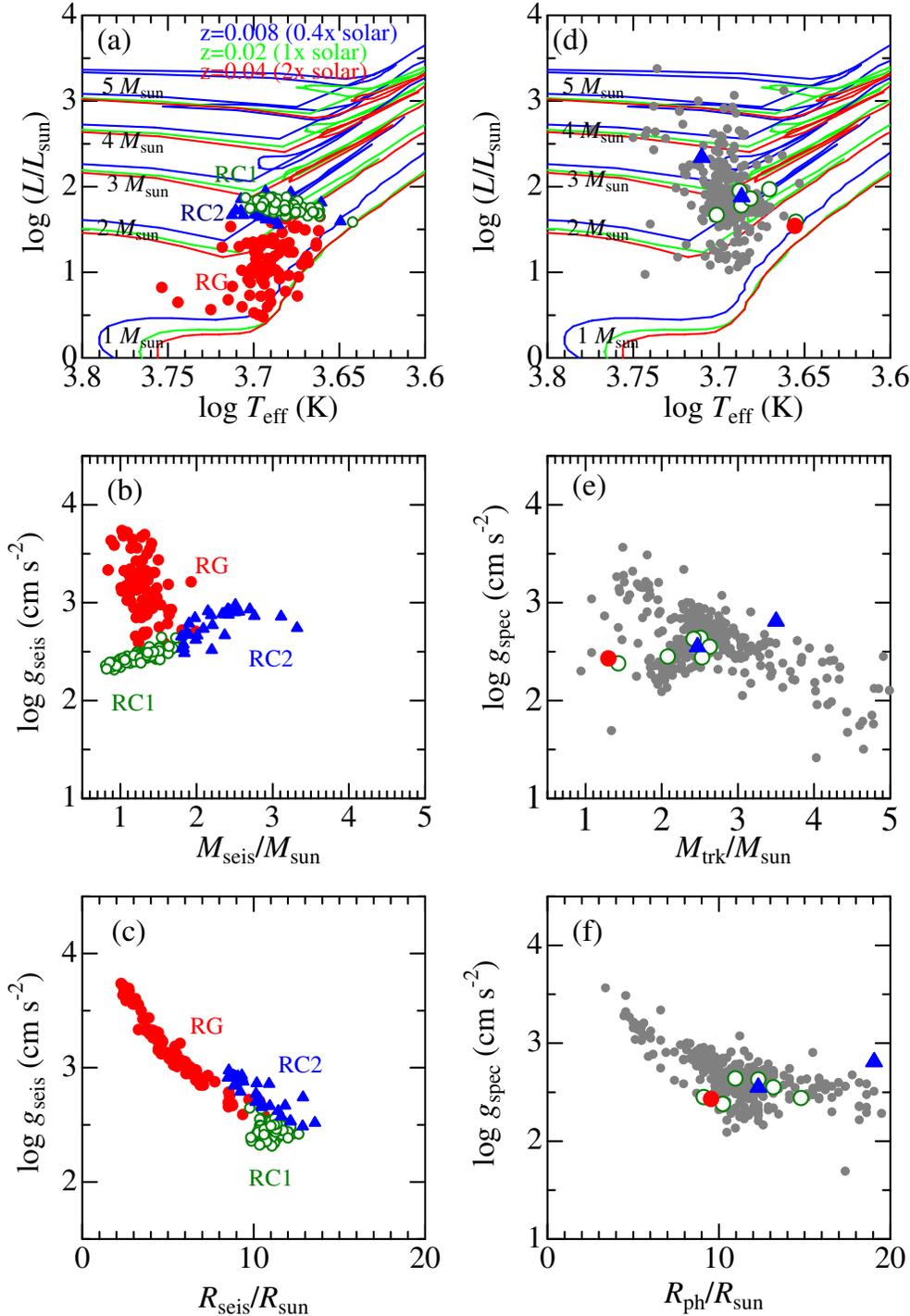}
\caption{Upper (a, d), middle (b, e), and lower (c, f) panels show the 
diagrams of $\log L$ vs. $\log T_{\rm eff}$ (with theoretical evolutionary 
tracks computed by Lejeune \& Schaerer 2001), $\log g$ vs. $M$, and
$\log g$ vs. $R$, respectively.
The left panels (a, b, c) are constructed from the seismic results of 218 
giants published by Mosser et al. (2012), where different symbols correspond 
to each evolutionary class: filled circles $\cdots$ normal red giants (RG),
open circles $\cdots$ 1st clump giants (RC1), and filled triangles $\cdots$ 
2nd clump giants. In panel (a), Brown et al.'s (2011) $T_{\rm eff}$(KIC) values 
were used for the abscissa as well as for evaluation of $L$. 
In the right panels (d, e, f) are plotted the results of 322 giants studied 
in Paper I [panels (d) and (e) here are same as Fig. 2 and Fig. 3e therein],
where $T_{\rm eff}$ and $\log g$ were spectroscopically determined,
$L$ and $R$ are the photometric values (cf. caption of Table 2),
and $M$ was estimated in comparison with evolutionary tracks. 
In addition, the results of 9 stars (among our 55 program stars) with known 
parallaxes (see Table 2) are also overplotted in panels (d), (e), and (f), 
which will be discussed in Sect. 4.2. 
}
\label{fig2}
\end{minipage}
\end{figure*}

As a probe, we compared the relations between the parameters ($\log L$ vs. 
$\log T_{\rm eff}$, $\log g$ vs. $M$, and $\log g$ vs. $R$) of 322 red giants 
studied in Paper I and those corresponding to Mosser et al.'s (2012) 218 samples 
in the $Kepler$ field (with well established parameters and evolutionary status), 
as depicted in Fig. 2.
Then, an interesting similarity (i.e., the characteristic ``trifid'' structure) 
is noticed between Fig. 2b and Fig. 2e, which suggests a possibility that 
$\log g$ and $M$ values determined in such an ordinary manner as in Paper I 
may be used to clarify (or at least guess) the evolutionary status of 
a red giant from its position on the $\log g$ vs. $M$ diagram. 
If this is confirmed to be practically feasible, it would be very useful.  

However, some concerns still exist. First, we do not yet have much confidence
as to whether the spectroscopic $\log g$ values determined in Paper I are 
sufficiently reliable, since they tend to be systematically lower 
by $\sim$~0.2--0.3 dex as compared to various literature values despite of 
being derived in a similar way (cf. Sect. 3.3 in Paper I; especially
Figs. 5b, 6b, 7b, and 8b therein). 
Second, the intersection of the ``trifid'' structure is located at 
$\sim$~1.5--2~M$_{\odot}$ for Mosser et al.'s (2012) sample (Fig. 2b), while 
at an appreciably larger mass around $\sim$~2.5~M$_{\odot}$ for those in Paper I 
(Fig. 2e). This difference makes us wonder whether the mass values derived 
from evolutionary tracks are really correct.

Thus, as a first preparatory step toward our intended goal, we decided to carry out 
a spectroscopic study for a number of red giant stars in the {\it Kepler} field 
(with well-established evolutionary stage and $M_{\rm seis}$/$R_{\rm seis}$/$\log g_{\rm seis}$) 
selected from Mosser et al.'s  (2012) sample. Based on the high-dispersion spectra 
primarily obtained with Subaru/HDS, their parameters were determined (while 
following the same manner as in Paper I) and compared with the seismic values, 
by which we may be able to check the accuracy level of our parameter determination 
procedures. This was the original motivation of the present study, and 
the purpose of this article is report the outcome of the investigation.

\section{Observational data}

The targets in this investigation were exclusively taken from Mosser et al.'s (2012)
218 red giants in the {\it Kepler} field, for which the evolutionary status
is well established in asteroseismic manner along with $R_{\rm seis}$ and $M_{\rm seis}$. 
The observations of 42 stars (with their {\it Kepler} magnitudes being $\sim$~9--11~mag) 
selected from this list were carried out on 2014 September 9 (UT) 
with the High Dispersion Spectrograph (HDS; Noguchi et al. 2002) placed at 
the Nasmyth platform of the 8.2-m Subaru Telescope, by which high-dispersion 
spectra covering $\sim$~5100--7800~$\rm\AA$ were obtained (with two CCDs of 
2K$\times$4K pixels) in the standard StdRa setting with the red cross disperser. 
We used the Image Slicer \#2 (Tajitsu, Aoki \& Yamamuro 2012), which resulted in a 
spectrum resolving power of $R \simeq 80000$. The total exposure time 
per star was from a few minutes to $\sim$~10~min in most cases. 

The reduction of the spectra (bias subtraction, flat-fielding, 
scattered-light subtraction, spectrum extraction, wavelength calibration,
co-adding of frames to improve S/N, continuum normalization) was 
performed by using the ``echelle'' package of 
the software IRAF\footnote{IRAF is distributed
    by the National Optical Astronomy Observatories,
    which is operated by the Association of Universities for Research
    in Astronomy, Inc. under cooperative agreement with
    the National Science Foundation.} 
in a standard manner. 
Typical S/N ratios attained in the final spectra are $\sim$~100--200.

In addition to these Subaru/HDS data, we also used the spectra of {\it Kepler} 
red giants published by Thygesen et al. (2012), which they obtained with 
Nordic Optical Telescope (NOT; $R \simeq 67000$, 3700--7300~$\rm\AA$, S/N$\simeq$~80--100), 
Canada-France-Hawaii Telescope (CFHT; $R \simeq 80000$, 3700--10500~$\rm\AA$, 
S/N$\simeq$~200), and Telescope Bernard Lyot (TBL; $R \simeq 75000$, 
3700--10500~$\rm\AA$, S/N$\simeq$~200). Since 16 stars (out of their 82 stars) 
are included in Mosser et al.'s (2012) list (218 stars), we used their 
spectra for these 16 stars. Note that the spectra for three stars (KIC~1726211, 
KIC~2714397, and KIC~3744043) are commonly available in both samples of ours 
as well as Thygesen et al.'s, which means that the net number of our targets 
is 55 ($=42+16-3$). The spectra for these three stars in the orange region 
(around $\sim 6085$~$\rm\AA$) are displayed in Fig. 3 for comparison. The complete 
target list and the distinction of data source [ours $\cdots$ ``Subaru'', 
Thygesen et al. $\cdots$ ``NOT'' or ``CFHT/TBL'')\footnote{We could not distinguish 
whether Thygesen et al.'s (2012) spectra covering wide wavelength region 
(3700--10500~$\rm\AA$) correspond to which of CFHT or TBL, since the details are not 
specified. Therefore, we simply noted as ``CFHT/TBL'' for the relevant cases.}] 
is given in Table 1. 

\setcounter{table}{0}
\begin{table*}
\begin{minipage}{180mm}
\scriptsize
\caption{Basic data and the resulting parameters of the program stars.}
\begin{center}
\begin{tabular}{crccccrrrrccccc}\hline
KIC\# & $Kepler$ & $T_{\rm eff}$ & $\log g$ & $v_{\rm t}$ & [Fe/H] & $\nu_{\rm max}$ &
$\Delta \nu$ & $\Delta\Pi_{1}$ & $R_{\rm seis}$ & $M_{\rm seis}$ & $\log g_{\rm seis}$ & 
$v_{\rm e}\sin i$ & class & Source \\
(1) & (2) & (3) & (4) & (5) & (6) & (7) & (8) & (9) & (10) & (11) & (12) &
(13) & (14) & (15) \\
\hline
\multicolumn{14}{c}{[Results based on our Subaru/HDS spectra]} \\
01726211& 10.93& 4983& 2.49& 1.34& $-$0.57&  29.0&  3.72&326.10& 11.62& 1.19&  2.39&  2.5&RC1&  Subaru \\
02013502& 11.91& 4913& 2.69& 1.14& $-$0.02&  61.0&  5.72&232.20& 10.26& 1.94&  2.71&  2.8&RC2&  Subaru \\
02303367& 10.18& 4601& 2.39& 1.36& +0.06&  34.2&  4.05&308.70& 11.11& 1.23&  2.44&  2.2&RC1&  Subaru \\
02424934& 10.38& 4792& 2.48& 1.31& $-$0.18&  33.0&  3.91&328.50& 11.73& 1.36&  2.43&  2.1&RC1&  Subaru \\
02448225& 10.81& 4577& 2.37& 1.32& +0.16&  38.2&  3.96&285.50& 12.94& 1.87&  2.49&  2.4&RC2&  Subaru \\
02714397& 10.51& 4910& 2.56& 1.38& $-$0.47&  33.0&  4.14&326.70& 10.59& 1.12&  2.44&  2.4&RC1&  Subaru \\
03217051& 11.38& 4590& 2.44& 1.27& +0.21&  36.0&  4.22&314.50& 10.75& 1.22&  2.46&  2.4&RC1&  Subaru \\
03455760& 10.94& 4654& 2.68& 1.13& $-$0.07&  48.0&  4.89& 64.30& 10.75& 1.63&  2.59&  2.1&RG&  Subaru \\
03730953&  8.82& 4861& 2.55& 0.97& $-$0.07&  50.3&  4.91&308.60& 11.42& 1.97&  2.62&  2.8&RC2&  Subaru \\
03744043&  9.66& 4946& 2.94& 1.10& $-$0.35& 110.9&  9.90& 75.98&  6.25& 1.31&  2.97&  1.6&RG&  Subaru \\
03758458& 11.28& 5009& 2.71& 1.28& +0.07&  65.0&  5.87&262.62& 10.48& 2.18&  2.74&  2.6&RC2&  Subaru \\
04036007& 11.43& 4916& 2.42& 1.33& $-$0.36&  38.0&  4.37&307.99& 10.96& 1.38&  2.50&  2.2&RC1&  Subaru \\
04044238&  7.83& 4519& 2.38& 1.32& +0.20&  33.0&  4.07&296.35& 10.52& 1.06&  2.42&  2.2&RC1&  Subaru \\
04243623& 11.17& 5005& 3.75& 0.74& $-$0.31& 495.6& 32.80&118.20&  2.56& 0.99&  3.62&  1.9&RG&  Subaru \\
04243796& 10.80& 4620& 2.34& 1.34& +0.11&  37.0&  4.28&296.27& 10.78& 1.26&  2.48&  2.1&RC1&  Subaru \\
04351319&  9.94& 4876& 3.32& 0.91& +0.29& 386.0& 24.50& 98.70&  3.53& 1.45&  3.51&  2.1&RG&  Subaru \\
04445711& 10.72& 4876& 2.50& 1.38& $-$0.32&  37.0&  4.29&333.30& 11.02& 1.35&  2.49&  2.3&RC1&  Subaru \\
04770846&  9.54& 4847& 2.60& 1.27& +0.02&  53.9&  5.46&308.12&  9.88& 1.58&  2.65&  2.5&RC1&  Subaru \\
04902641& 10.76& 4987& 2.84& 1.15& +0.03& 101.6&  7.87&147.28&  9.10& 2.56&  2.93&  2.2&RC2&  Subaru \\
04952717& 11.23& 4793& 3.13& 0.93& +0.13& 202.1& 15.58& 87.52&  4.53& 1.24&  3.22&  1.6&RG&  Subaru \\
05000307& 11.23& 5023& 2.64& 1.27& $-$0.25&  42.2&  4.74&323.70& 10.45& 1.41&  2.55&  2.9&RC1&  Subaru \\
05033245& 10.91& 5049& 3.41& 0.97& +0.11& 424.2& 26.82&114.20&  3.29& 1.41&  3.55&  2.1&RG&  Subaru \\
05088362& 11.15& 4760& 2.41& 1.33& +0.03&  40.1&  3.99&230.95& 13.65& 2.22&  2.52&  2.1&RC2&  Subaru \\
05128171& 10.10& 4808& 2.54& 1.30& +0.04&  56.1&  5.27&319.59& 11.00& 2.03&  2.66&  2.4&RC2&  Subaru \\
05266416& 10.61& 4767& 2.50& 1.33& $-$0.09&  32.4&  3.75&284.37& 12.49& 1.51&  2.42&  2.3&RC1&  Subaru \\
05307747&  8.40& 5031& 2.77& 1.27& +0.01&  88.2&  6.88&296.30& 10.38& 2.91&  2.87&  2.6&RC2&  Subaru \\
05530598&  8.72& 4599& 2.85& 1.04& +0.37& 105.5&  8.72& 76.30&  7.39& 1.68&  2.93&  2.1&RG&  Subaru \\
05723165& 10.60& 5255& 3.67& 0.90& $-$0.02& 578.8& 34.67&127.30&  2.74& 1.36&  3.70&  2.2&RG&  Subaru \\
05737655&  7.20& 5026& 2.45& 1.46& $-$0.63&  29.8&  4.24&278.60&  9.23& 0.78&  2.40&  4.0&RC1&  Subaru \\
05806522& 11.26& 4574& 2.60& 1.09& +0.12&  56.0&  5.92& 69.47&  8.49& 1.18&  2.65&  1.8&RG&  Subaru \\
05866737& 10.76& 4874& 2.86& 1.14& $-$0.26&  67.6&  6.55& 59.51&  8.64& 1.52&  2.75&  1.8&RG&  Subaru \\
05990753& 10.92& 5011& 2.97& 1.15& +0.19&  97.9&  7.57&195.00&  9.50& 2.70&  2.92&  3.4&RC2&  Subaru \\
06117517& 10.59& 4649& 2.94& 0.97& +0.28& 116.9& 10.16& 76.91&  6.06& 1.26&  2.98&  1.8&RG&  Subaru \\
06144777& 10.69& 4734& 3.02& 1.02& +0.14& 126.0& 11.01& 69.91&  5.62& 1.18&  3.01&  1.9&RG&  Subaru \\
06276948& 10.83& 4939& 2.84& 1.18& +0.19&  91.6&  7.41&259.00&  9.21& 2.35&  2.88&  2.6&RC2&  Subaru \\
07205067& 10.72& 5064& 2.58& 1.32& +0.03&  66.0&  5.30&278.60& 13.13& 3.49&  2.75&  3.8&RC2&  Subaru \\
07581399& 11.49& 5070& 2.74& 1.14& +0.01&  85.0&  6.59&222.00& 10.94& 3.13&  2.86&  3.3&RC2&  Subaru \\
08378462& 11.13& 4995& 2.81& 1.19& +0.06&  90.3&  7.27&238.30&  9.48& 2.47&  2.88&  2.1&RC2&  Subaru \\
08702606&  9.29& 5472& 3.66& 1.01& $-$0.11& 630.3& 39.70&177.00&  2.32& 1.09&  3.74&  2.7&RG&  Subaru \\
08718745& 10.71& 4902& 3.20& 0.98& $-$0.25& 129.4& 11.40& 79.45&  5.47& 1.17&  3.03&  1.8&RG&  Subaru \\
09173371&  9.27& 5064& 2.85& 1.18& +0.00&  98.9&  7.95&236.34&  8.74& 2.32&  2.92&  2.9&RC2&  Subaru \\
11819760& 10.80& 4824& 2.36& 1.32& $-$0.18&  29.1&  3.64&296.20& 11.98& 1.25&  2.38&  1.8&RC1&  Subaru \\
\hline
\multicolumn{15}{c}{[Results based on Thygesen et al.'s (2012) spectra]}\\
01726211& 10.93& 4933& 2.37& 1.33& $-$0.57&  29.0&  3.72&326.10& 11.56& 1.17&  2.38&  2.3&RC1& NOT \\
02714397& 10.51& 4956& 2.77& 1.39& $-$0.36&  33.0&  4.14&326.70& 10.64& 1.14&  2.44&  2.0&RC1& NOT \\
03744043&  9.66& 4938& 2.88& 1.05& $-$0.28& 110.9&  9.90& 75.98&  6.24& 1.31&  2.97&  1.6&RG& NOT \\
03748691& 11.59& 4762& 2.53& 1.31& +0.11&  37.0&  4.24&299.60& 11.15& 1.37&  2.48&  2.2&RC1& NOT \\
05795626&  9.16& 4923& 2.38& 1.30& $-$0.72&  37.2&  4.45&311.10& 10.35& 1.21&  2.49&  2.5&RC1& NOT \\
06690139& 11.67& 4979& 3.02& 1.06& $-$0.14& 113.8&  9.69& 66.90&  6.72& 1.56&  2.98&  1.9&RG& NOT \\
08813946&  7.03& 4862& 2.64& 1.27& +0.09&  72.8&  6.39&255.40&  9.76& 2.09&  2.78&  3.0&RC1& CFHT/TBL \\
09705687&  9.58& 5129& 2.81& 1.26& $-$0.19&  72.3&  6.62&255.90&  9.28& 1.92&  2.79&  2.5&RC2& NOT \\
10323222&  6.72& 4525& 2.43& 1.18& +0.04&  47.7&  4.88& 61.25& 10.58& 1.55&  2.58&  2.0&RG& CFHT/TBL \\
10404994&  7.42& 4803& 2.63& 1.26& $-$0.06&  40.3&  4.43&297.00& 11.17& 1.50&  2.52&  2.6&RC1& CFHT/TBL \\
10426854&  9.94& 4968& 2.52& 1.34& $-$0.30&  40.9&  4.35&318.00& 11.96& 1.78&  2.53&  2.6&RC1& NOT \\
10716853&  6.89& 4874& 2.55& 1.27& $-$0.08&  48.8&  4.95&296.00& 10.92& 1.75&  2.61&  2.2&RC1& CFHT/TBL \\
11444313& 11.31& 4757& 2.52& 1.36& +0.00&  34.2&  3.98&315.20& 11.69& 1.39&  2.45&  2.4&RC1& NOT \\
11569659& 11.51& 4879& 2.48& 1.36& $-$0.26&  29.2&  4.04&292.60&  9.81& 0.85&  2.38&  2.6&RC1& NOT \\
11657684& 11.71& 4951& 2.47& 1.29& $-$0.12&  33.7&  4.09&291.30& 11.13& 1.27&  2.45&  1.9&RC1& NOT \\
12884274&  7.59& 4683& 2.44& 1.35& +0.11&  41.4&  4.57&323.20& 10.65& 1.39&  2.53&  2.4&RC1& CFHT/TBL \\
\hline
\end{tabular}
\end{center}
Following the serial number and the $Kepler$ magnitude (in mag) of the {\it Kepler} Input Catalogue 
(KIC; cf. Brown et al. 2011) in Columns (1) and (2), the atmospheric parameters
(effective temperature $T_{\rm eff}$ in K, logarithmic surface gravity $\log g$ in 
cm~s$^{-2}$/dex, microturbulent velocity dispersion $v_{\rm t}$ in km~s$^{-1}$,
and metallicity [Fe/H] in dex) spectroscopically determined from Fe~{\sc i} and 
Fe~{\sc ii} lines  are presented in Columns (3)--(6).
Columns (7)--(9) give the asteroseismic quantities (expressed in $\mu$Hz) taken from 
Mosser et al. (2012): the central frequency of the oscillation power excess ($\nu_{\rm max}$), 
the large frequency separation ($\Delta\nu$), and the gravity-mode spacing ($\Delta\Pi_{1}$;
good indicator for discriminating between RG and RC1/RC2). 
Presented in Columns (10)--(12) are the seismic radius (in R$_{\odot}$), seismic mass (in M$_{\odot}$), 
and the corresponding seismic surface gravity (in cm~s$^{-2}$/dex, which were evaluated 
from $\nu_{\rm max}$ and $\Delta\nu$ by using the scaling relations [Eqs. (1)--(3)].
In Columns (13) and (14) are given the projected rotational velocity (in km~s$^{-1}$;
derived from the 6080--6089~$\rm\AA$ fitting analysis) and the evolutionary class
determined by Mosser et al. (2012) (RG: red giant, RC1: 1st clump giant, RC2: 2nd clump giant).
Column (15) gives the data source (cf. Sect. 2 and footnote 2) 
of the spectrum.  
\end{minipage}
\end{table*}

\setcounter{figure}{2}
\begin{figure}
\begin{minipage}{80mm}
\includegraphics[width=8.0cm]{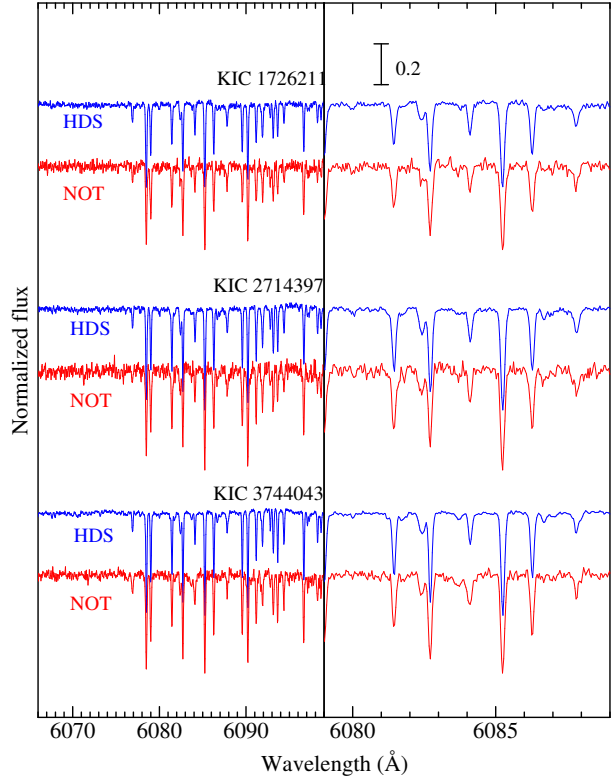}
\caption{Comparison of the representative spectra used in this study.
Shown are the spectral portions in the 6066--6099~$\rm\AA$ region
(left panel) and the 6079--6089~$\rm\AA$ region (right panel)
for KIC~1726211, KIC~2714397, and KIC~3744043,
for which Subaru/HDS spectra (based on our observation) 
and NOT spectra (published by Thygesen et al. 2012) are
doubly available. Each spectrum is appropriately shifted
relative to the adjacent one. The wavelength scale is adjusted to the
laboratory frame. 
}
\label{fig3}
\end{minipage}
\end{figure}

\section{Parameter determinations}

\subsection{Atmospheric parameters}

The determination of the atmospheric parameters ($T_{\rm eff}$, $\log g$, 
$v_{\rm t}$ and [Fe/H]) was implemented in the same manner as in Paper I
(see Sect. 3.1 for the details) by applying the program TGVIT (Takeda et al. 2005;
cf. Sect. 2 therein), which is based on the principles described in 
Takeda, Ohkubo \& Sadakane (2002), to the equivalent widths ($W_{\lambda}$) of 
Fe I and Fe II lines measured on the spectrum of each star by the Gaussian-fitting method. 
As before, we restricted to using lines satisfying $W_{\lambda} \le 120$~m$\rm\AA$ 
and those showing abundance deviations from the mean larger than $2.5 \sigma$ 
were rejected. The final number of adopted lines are $\sim 200$ (Fe~{\sc i}) 
and $\sim 15$ (Fe~{\sc i}) on the average.
The resulting final solutions are presented in Table 1, while $\log\epsilon$(Fe) 
values (Fe abundances corresponding to the final solutions) are plotted 
against $W_{\lambda}$ and $\chi_{\rm low}$ in Fig. 4 and Fig. 5,
respectively, where we can see that the required conditions 
[i.e., no systematic dependence of $\log\epsilon$(Fe) upon $W_{\lambda}$
as well as $\chi_{\rm low}$, and the equality of 
$\langle\log\epsilon$(Fe~{\sc i})$\rangle$ = 
$\langle\log\epsilon$(Fe~{\sc ii})$\rangle$ (ionization equilibrium)] 
are reasonably accomplished.
The internal statistical errors involved with these solutions
are almost the same order as the case of Paper I.
The detailed $W_{\lambda}$ and $\log\epsilon$(Fe) data for each star are given 
in tableE1.dat (supplementary online material).

\setcounter{figure}{3}
\begin{figure*}
\begin{minipage}{160mm}
\includegraphics[width=16.0cm]{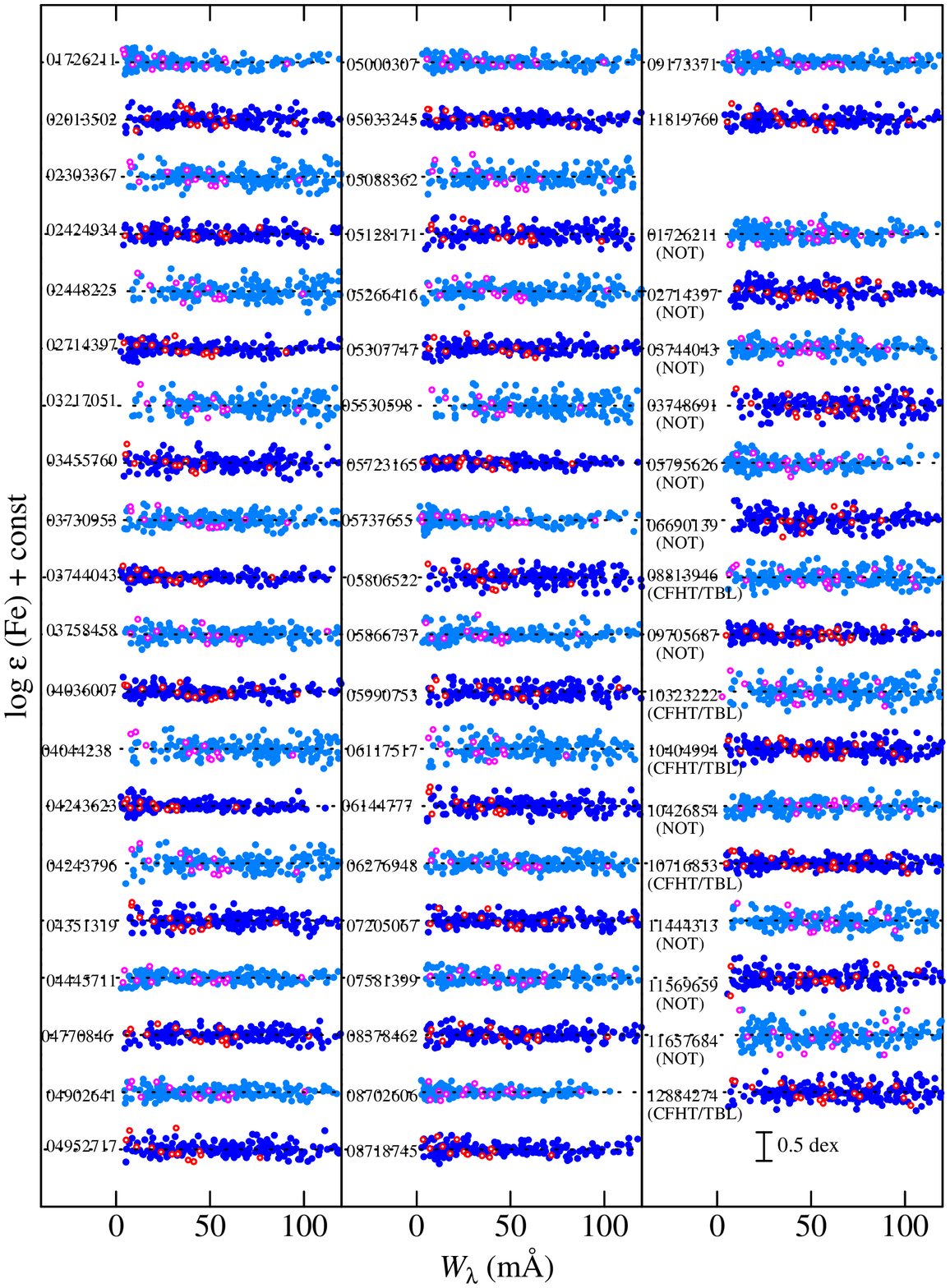}
\caption{Fe abundance vs. equivalent width relations 
corresponding to the finally established atmospheric parameters of 
$T_{\rm eff}$, $\log g$, and $v_{\rm t}$ for each of the 58 stars,
being arranged in the same order as in Table 1 as indicated by the 
KIC number.
The filled and open symbols correspond to Fe~{\sc i} and Fe~{\sc ii} 
lines, respectively. The results for each star are shown relative to 
the mean abundance indicated by the horizontal dotted line, and 
vertically shifted by 1.0 relative to the adjacent ones.
}
\label{fig4}
\end{minipage}
\end{figure*}

\setcounter{figure}{4}
\begin{figure*}
\begin{minipage}{160mm}
\includegraphics[width=16.0cm]{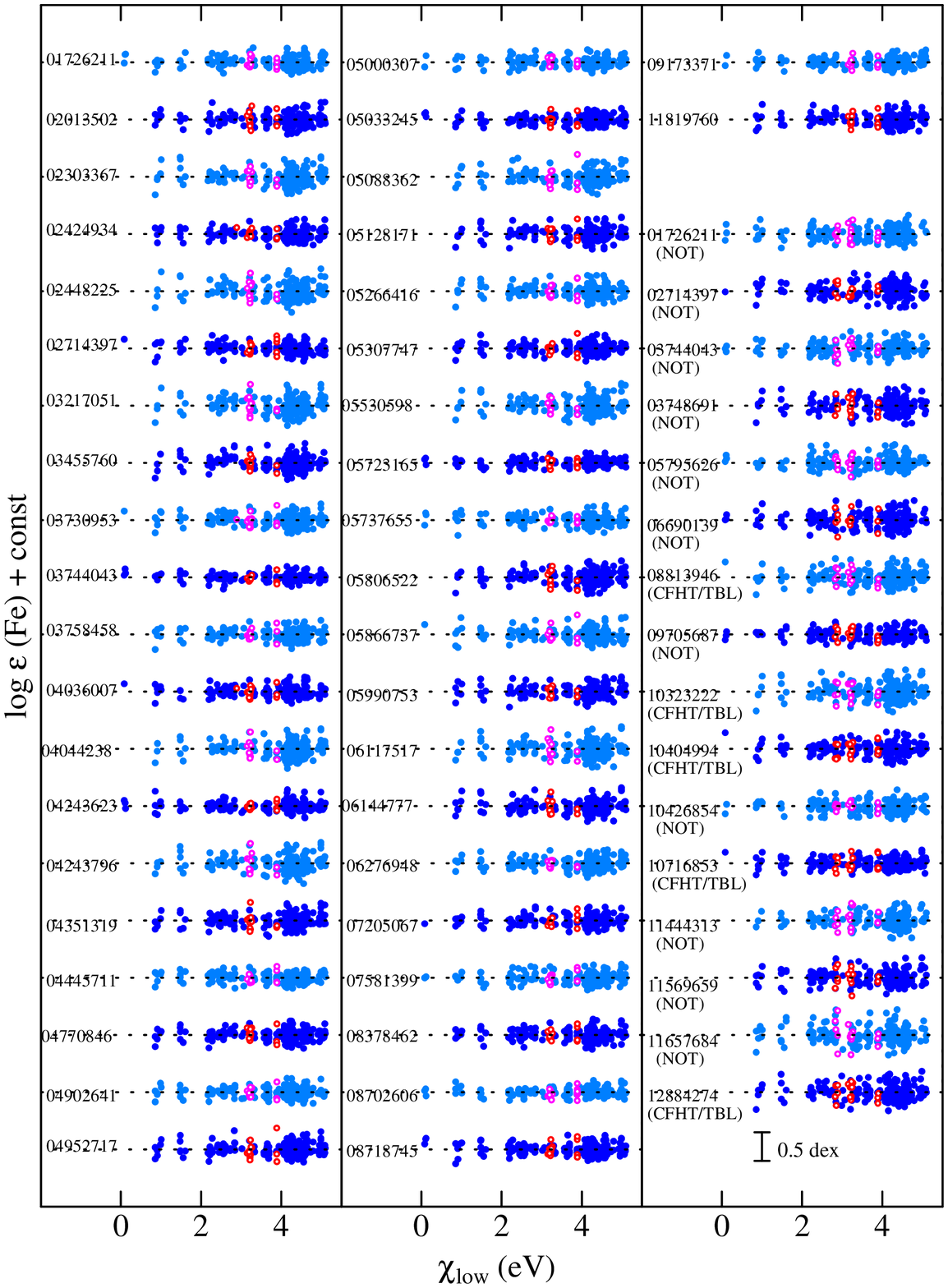}
\caption{Fe abundance vs. lower excitation potential relations 
corresponding to the finally established atmospheric parameters 
for each of the 58 stars. Otherwise, the same as in Fig. 4.
}
\label{fig5}
\end{minipage}
\end{figure*}

\subsection{Rotational velocity}

Given such established atmospheric parameters, we constructed the model atmospheres
for each of the program stars as done in Paper I.   
Then, based on these atmospheric models, the projected rotational velocity 
($v_{\rm e}\sin i$) of each star was determined from the width of the macrobroadening 
function (consisting of combined effects of instrumental broadening, macroturbulence 
broadening, and rotational broadening), which was derived from the spectrum-fitting 
in the 6080--6089~$\rm\AA$ region in the same way as done in Paper I (cf. Sect. 4.2 
therein for more details). Fig. 6 shows how the best-fit theoretical spectrum matches
the observed spectrum for each star.
The resulting $v_{\rm e}\sin i$ values are presented in Table 1.
Besides, the detailed values of the relevant broadening widths (the instrumental width
and the macroturbulence width estimated from $\log g$, which are to be subtracted from
the total broadening) are given in tableE2.dat (supplementary online material), 
where the elemental abundances of Si, Ti, V, Fe, Co, and Ni obtained as by-products
are also given.\footnote{Since chemical abundances are outside of the scope of this 
paper, we refrain from discussing these results further on. We only note that 
the resulting [X/Fe] vs. [Fe/H] relations for these elements (in the metallicity 
range of $-0.7 \la$~[Fe/H]~$\la +0.3$) are quite similar to Figs. 11r--11t in Paper I.}

\setcounter{figure}{5}
\begin{figure*}
\begin{minipage}{160mm}
\includegraphics[width=16.0cm]{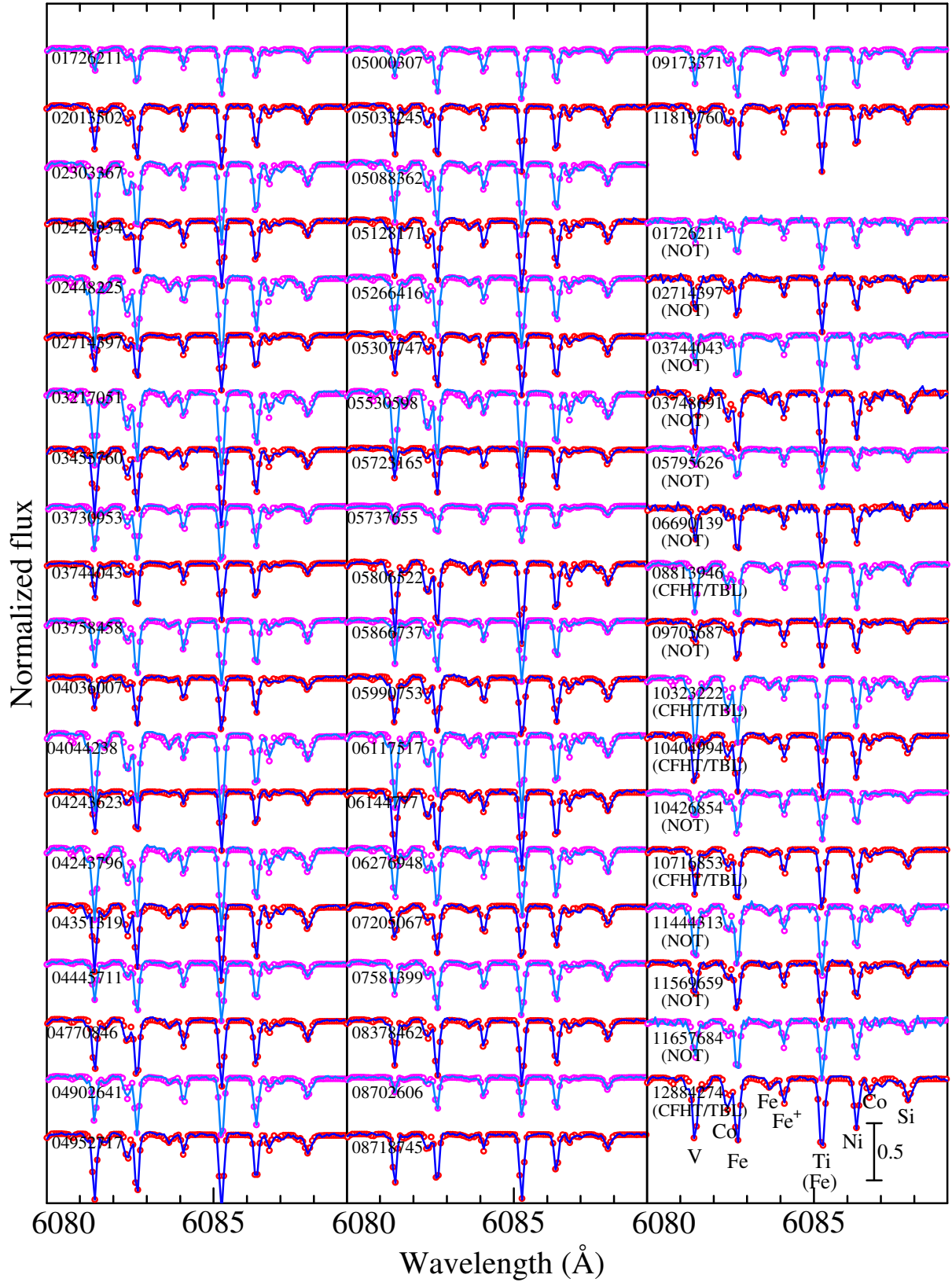}
\caption{Synthetic spectrum fitting in the 6080--6089~$\rm\AA$
region accomplished by varying the abundances of 
Si, Ti, V, Fe, Co, and Ni, along with the macrobroadening 
velocity ($v_{\rm M}$) and the wavelength shift (radial velocity).
The best-fit theoretical spectra are shown by solid lines, 
while the observed data are plotted by symbols, where
the wavelength scale of the stellar spectrum has been adjusted
to the laboratory frame.  
In each panel, the spectra are arranged in the same order as in Table 1 
(corresponding KIC numbers are indicated), and each spectrum is vertically 
shifted by 0.5 relative to the adjacent one. 
}
\label{fig6}
\end{minipage}
\end{figure*}

\subsection{Parameters related to HR diagram}

Unfortunately, distances are not necessarily known for all our 55 targets, 
in which Hipparcos parallax data (ESA 1997) are available for only 9 
comparatively bright stars (KIC 03730953, 04044238, 05737655, 08813946, 09705687, 
10323222, 10404994, 10716853, 12884274).  
We derived various stellar parameters (luminosity, radius, and mass) relevant to 
the HR diagram for these stars with the help of Lejeune \& Schaerer's (2001) 
theoretical evolutionary tracks in the same manner as done in Paper I 
(cf. Sect. 3.2 therein), which are summarized in Table 2.
Such derived quantities for these 9 stars are also overplotted in Figs. 2d--2f.
Note that the results for KIC~09705687 (HIP~94976) are considered
to be less reliable, since its parallax contains a large error (larger than 
the parallax value itself).

\setcounter{table}{1}
\begin{table*}
\begin{minipage}{180mm}
\scriptsize
\caption{Parameters derived from evolutionary tracks for 9 stars with available parallaxes.}
\begin{center}
\begin{tabular}{ccccccrcccccrcrcl}\hline
KIC\# & HIP\# & $V$ & $\pi$ & $\sigma_{\pi}$ & $A_{V}$ & $M_{V}$ & B.C. & $\log L$ & $T_{\rm eff}$ & 
[Fe/H] & $M_{\rm trk}$ & $R_{\rm ph}$ & $M_{\rm seis}$ & $R_{\rm seis}$ & class & Remark \\ 
(1) & (2) & (3) & (4) & (5) & (6) & (7) & (8) & (9) & (10) & (11) & (12) & (13) & (14) & (15) & (16) & (17) \\
\hline
03730953&  93687&  9.07&  2.06& 0.95&  0.27&  0.37& $-$0.31&  1.88&  4861& $-$0.07&  2.47& 12.3&  1.97& 11.4&RC2& \\
04044238&  94221&  8.15&  4.36& 0.78&  0.10&  1.25& $-$0.48&  1.59&  4519& +0.20&  1.43& 10.2&  1.06& 10.5&RC1& \\
05737655&  98269&  7.32&  5.22& 0.61&  0.08&  0.83& $-$0.26&  1.67&  5026& $-$0.63&  2.08&  9.1&  0.78&  9.2&RC1& \\
08813946&  95005&  7.19&  5.04& 0.59&  0.09&  0.61& $-$0.31&  1.78&  4862& +0.09&  2.51& 11.0&  2.09&  9.8&RC1& \\
09705687&  94976&  9.79&  0.82& 1.08&  0.24& $-$0.88& $-$0.22&  2.34&  5129& $-$0.20&  3.50& 19.1&  1.92&  9.3&RC2& unreliable $\pi$ \\
10323222&  92885&  7.01&  7.73& 0.61&  0.08&  1.37& $-$0.48&  1.54&  4525& +0.04&  1.30&  9.5&  1.55& 10.6&RG& \\
10404994&  95687&  7.67&  3.76& 0.65&  0.11&  0.44& $-$0.33&  1.86&  4803& $-$0.06&  2.42& 12.3&  1.50& 11.2&RC1& \\
10716853&  93376&  7.04&  4.48& 0.52&  0.11&  0.19& $-$0.31&  1.95&  4874& $-$0.08&  2.63& 13.2&  1.75& 10.9&RC1& \\
12884274&  94896&  7.88&  3.04& 0.69&  0.09&  0.20& $-$0.39&  1.97&  4683& +0.11&  2.53& 14.8&  1.39& 10.6&RC1& \\
\hline
\end{tabular}
\end{center}
Column (1) --- KIC number,
Column (2) --- Hipparcos number,
Column (3) --- apparent visual magnitude (in mag),
Column (4) --- parallax (in milliarcsec), 
Column (5) --- error of $\pi$ (in milliarcsec),
Column (6) --- interstellar extinction (in mag),
Column (7) --- absolute visual magnitude (in mag),
Column (8) --- bolometric correction (in mag),
Column (9) --- log(bolometric luminosity) (in L$_{\odot}$),
Column (10) --- spectroscopically determined $T_{\rm eff}$,
Column (11) --- spectroscopically determined [Fe/H],
Column (12) --- stellar mass (in M$_{\odot}$) estimated from theoretical evolutionary tracks,
Column (13) --- photometric radius (in R$_{\odot}$) derived from $L$ and $T_{\rm eff}$,
Column (14) --- seismic mass (in M$_{\odot}$),
Column (15) --- seismic radius (in R$_{\odot}$), and
Column (16) --- evolutionary status (see the caption of Table 1).  
The data in Columns (2)--(5) were taken from the Hipparcos catalogue (ESA 1997).
See Sect. 3.2 in Paper I for more details regarding the derivation of $A_{V}$, B.C., and $M_{\rm trk}$. 
Note that $T_{\rm eff}$, [Fe/H], $M_{\rm seis}$, and $R_{\rm seis}$ are the same as in Table 1. 
\end{minipage}
\end{table*}

\subsection{Comparison with previous studies}

Fig. 7 shows the comparison of our spectroscopic $T_{\rm eff}$, $\log g$, 
and [Fe/H] with those given in the {\it Kepler} Input Catalogue (Brown et al. 2011) 
which were estimated in the photometric way using color indices. 
We can state that almost the same conclusion as reported by Thygesen et al. (2012)
can be drawn from this comparison; i.e., a rough consistency can be seen 
(i.e., the correlation coefficient is $r \sim 0.7$; cf. Table 3) 
but very large discrepancies are sometimes observed, and the differences tend to be weakly   
$T_{\rm eff}$-dependent.\footnote{Note that the sign of the abscissa 
($T_{\rm eff}$) in our Figs. 7d--7f is inverse as compared with 
Thygesen et al.'s (2012) Fig. 2.}

\setcounter{table}{2}
\begin{table*}
\begin{minipage}{180mm}
\small
\caption{Statistical quantities for the stellar-parameter correlations shown in figures 7, 8, 11, and 12.}
\begin{center}
\begin{tabular}{ccccccccc}\hline
Figure & $X$ & $Y$ & $N$ & $r$ & 
$\langle Y - X \rangle$ & $\sigma$ & Unit & Remark \\ 
\hline
Fig. 7a & $T_{\rm eff}^{\rm spec}$ & $T_{\rm eff}^{\rm KIC}$ & 51 & 0.694 & $-15$ & 146 & K & \\
Fig. 7b & $\log g^{\rm spec}$ & $\log g^{\rm KIC}$ & 51 & 0.739 & $-0.02$ & 0.30 & dex & \\
Fig. 7c & [Fe/H]$^{\rm spec}$ & [Fe/H]$^{\rm KIC}$ & 51 & 0.668 & $-0.13$ & 0.26 & dex & \\
\hline
Fig. 8a & $T_{\rm eff}^{\rm ous}$ & $T_{\rm eff}^{\rm Thygesen}$ & 16 & 0.868 & $+40$ & 70 & K & \\
Fig. 8b & $\log g^{\rm ours}$ & $\log g^{\rm Thygesen}$ & 16 & 0.651 & $+0.01$ & 0.20 & dex & \\
Fig. 8c & $v_{\rm t}^{\rm ours}$ & $v_{\rm t}^{\rm Thygesen}$ & 16 & 0.905 & $+0.13$ & 0.04 & km~s$^{-1}$ & \\
Fig. 8d & [Fe/H]$^{\rm ours}$ & [Fe/H]$^{\rm Thygesen}$ & 16 & 0.965 & $-0.02$ & 0.08 & dex & \\
Fig. 8e & $v_{\rm e}\sin i^{\rm ours}$ & $v_{\rm e}\sin i^{\rm Thygesen}$ & 16 & 0.250 & $+0.89$ & 1.23 & km~s$^{-1}$ & \\
\hline
Fig. 11a & $\log g_{\rm seis}$ & $\log g_{\rm spec}$ & 58 & 0.960 & $-0.02$ & 0.10 & dex & \\
\hline
Fig. 12a & $\log M_{\rm seis}$ & $\log M_{\rm trk}$ & 8 & 0.389 & $+0.16$ & 0.15 & dex & KIC~9705687 excluded \\
Fig. 12b & $\log R_{\rm seis}$ & $\log R_{\rm ph}$ & 8 & 0.615 & $+0.04$ & 0.06 & dex & KIC~9705687 excluded \\
\hline
\end{tabular}
\end{center}
$X$ and $Y$ denote the quantities in the abscissa and ordinate of the relevant figure, respectively. 
while $N$ is the number of samples used for calculating the statistical quantities: $r$ is the correlation 
coefficient between $X$ and $Y$, $\langle Y - X \rangle$ is the average of the difference ($Y-X$), 
and $\sigma$ is the standard deviation. Note that logarithmic representation is used here for 
$M$ and $R$ in the last two rows, despite that these quantities are normally (i.e., linearly) 
represented in Fig. 12a and Fig. 12b. 
\end{minipage}
\end{table*}

\setcounter{figure}{6}
\begin{figure*}
\begin{minipage}{140mm}
\begin{center}
\includegraphics[width=14.0cm]{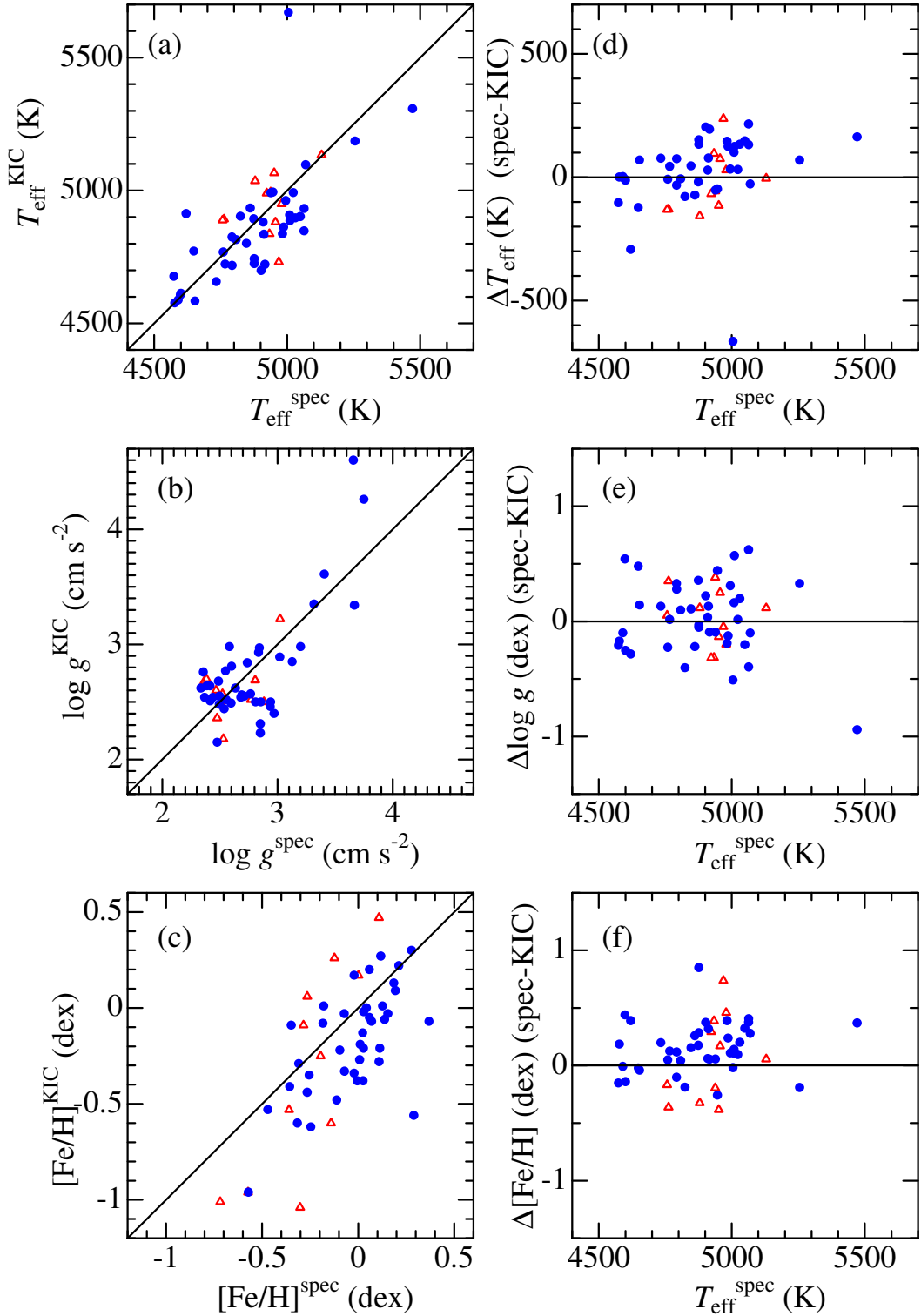}
\caption{
Comparison of $T_{\rm eff}$, $\log g$, and [Fe/H] given in
the {\it Kepler} Input Catalog (Brown et al. 2011) with those
we determined spectroscopically in this study.
Left panels (a, b, c) are the direct comparisons between these two,
while right panels (d, e, f) show the (ours$-$KIC) difference 
plotted against our spectroscopic $T_{\rm eff}$.
Filled circles are the results (for 42 stars) based on our Subaru/HDS spectra, 
while open triangles are those (for 16 stars) derived by using the NOT/CFHT/TBL 
spectra published by Thygesen et al. (2012).
 }
\label{fig7}
\end{center}
\end{minipage}
\end{figure*}

\setcounter{figure}{7}
\begin{figure*}
\begin{minipage}{140mm}
\begin{center}
\includegraphics[width=14.0cm]{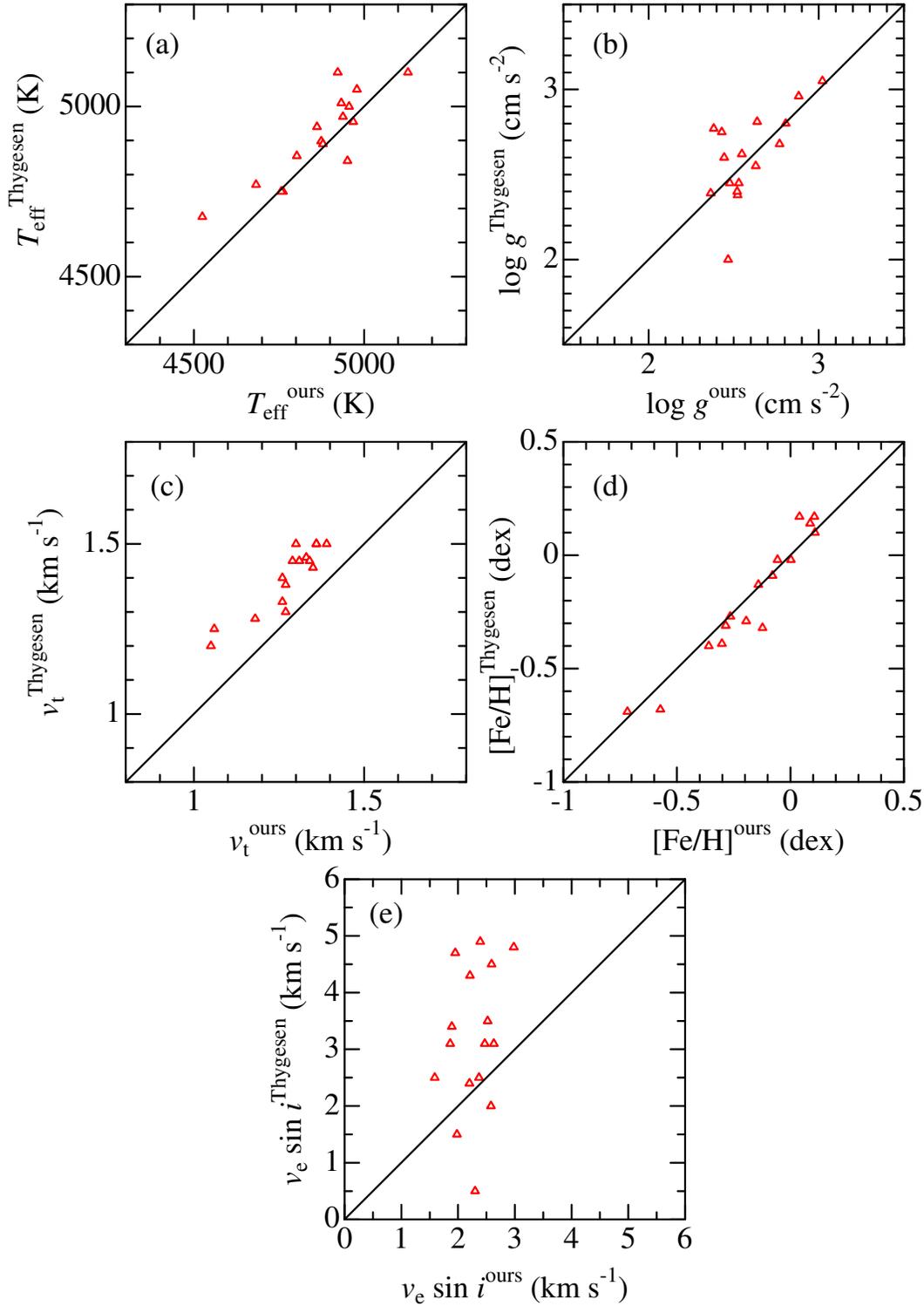}
\caption{Comparison of our (a) $T_{\rm eff}$, (b) $\log g$, (c) $v_{\rm t}$, 
(d) [Fe/H], and (e) $v_{\rm e} \sin i$ for 16 stars spectroscopically established 
by using Thygesen et al's (2012) spectra with their values similarly determined 
based on the same data. The meaning of the symbol is the same as in Fig. 7.
}
\label{fig8}
\end{center}
\end{minipage}
\end{figure*}

\setcounter{figure}{8}
\begin{figure*}
\begin{minipage}{150mm}
\begin{center}
\includegraphics[width=15.0cm]{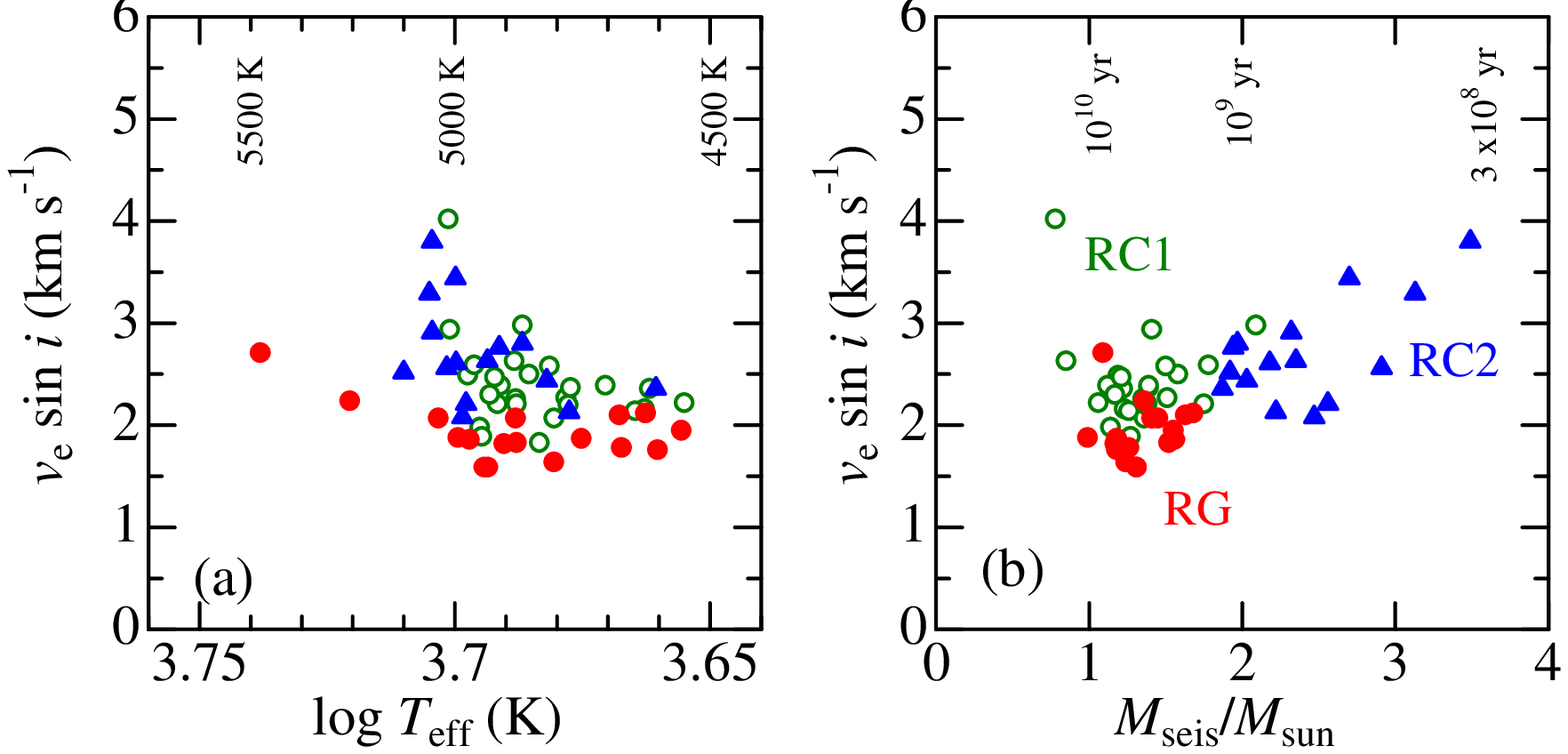}
\caption{The $v_{\rm e} \sin i$ values derived from the 6080--6089~$\rm\AA$
fitting plotted against $T_{\rm eff}$ (left panel) and $M_{\rm seis}$
(right panel) based on the data given in Table 1. 
See the caption of Fig. 2 for the meanings of the symbols.}
\label{fig9}
\end{center}
\end{minipage}
\end{figure*}

It is worth comparing our established parameters with those determined by 
Thygesen et al. (2012), since they are based on the same observational 
material as well as similar spectroscopic techniques. Such comparisons
of $T_{\rm eff}$, $\log g$, $v_{\rm t}$, [Fe/H], and $v_{\rm e} \sin i$ 
for 16 stars are displayed in Fig. 8.  We can see that both are in 
more or less reasonable agreement for $T_{\rm eff}$, $\log g$, $v_{\rm t}$, and [Fe/H] 
($r \sim$~0.7--0.9; cf. Table 3), 
though our $v_{\rm t}$ tends to be systematically smaller by 
$\sim$~0.1--0.2~km~s$^{-1}$ and our $T_{\rm eff}$ is slightly lower 
by $\sim 100$~K. However, our $v_{\rm e} \sin i$ values are in distinct
disagreement with no meaningful correlation ($r = 0.25$) 
with those derived by Thygesen et al. (2012), in the sense 
that ours show only a small dispersion ($\sim$~1.5--3~km~s$^{-1}$)
while theirs differ widely from each other ($\sim$~0.5--5~km~s$^{-1}$).
Since we found that their macroturbulence velocity (which they denoted 
as $v_{\rm macro}$ in their Table A.1) does not show any $\log g$-dependence,
which is in marked contrast with our basic postulation of $g$-dependent 
macroturbulence [cf. Eq.(1) in Paper I], the difference in the treatment 
of macroturbulence might be the cause for this large discrepancy.
For reference, our $v_{\rm e}\sin i$ results are plotted against 
$T_{\rm eff}$ and $M_{\rm seis}$ in Figs. 9a and 9b (in analogy with
Figs. 10e and 10f in Paper I), respectively, As we can see from these
figures, the trends of $v_{\rm e}\sin i$ in terms of these parameters are 
essentially the same as in Paper I (i.e., slowing-down tendency with a 
decrease in $T_{\rm eff}$, higher $v_{\rm e}\sin i$ with an increase in $M$).

\section{Discussion}

\subsection{Accuracy of spectroscopic gravity}

We first examine how our spectroscopically determined surface gravity is 
compared with the seismic gravity derived from $R_{\rm seis}$ and $M_{\rm seis}$.
Since the original $R_{\rm seis}$ and $M_{\rm seis}$ values presented by 
Mosser et al. (2012) correspond to Brown et al.'s (2011) photometric 
$T_{\rm eff}$(KIC), some of which appear to show considerable errors 
(cf.  Figs. 7a and 7d), we recalculated $R_{\rm seis}$ and $M_{\rm seis}$
by using our spectroscopic $T_{\rm eff}$ along with the seismic quantities 
[$\nu_{\rm max}$ (central frequency of the oscillation power excess) 
and $\Delta \nu$ (large frequency separation)] derived by Mosser et al. (2012).
For this purpose, we applied the scaling relations used by Kallinger et al. (2010a):
\begin{equation}
R_{\rm seis}/{\rm R}_{\odot} = (\nu_{\rm max}/\nu_{{\rm max},\odot}) 
(\Delta\nu/\Delta\nu_{\odot})^{-2} 
(T_{\rm eff}/{\rm T}_{{\rm eff},\odot})^{1/2}
\end{equation}
\begin{equation}
M_{\rm seis}/{\rm M}_{\odot} = (\nu_{\rm max}/\nu_{{\rm max},\odot}) ^{3}
(\Delta\nu/\Delta\nu_{\odot})^{-4} 
(T_{\rm eff}/{\rm T}_{{\rm eff},\odot})^{3/2}
\end{equation}
and
\begin{eqnarray}
g_{\rm seis}/{\rm g}_{\odot} = (M_{\rm seis}/{\rm M}_{\odot})
 (R_{\rm seis}/{\rm R}_{\odot})^{-2} \nonumber \\
 = (\nu_{\rm max}/\nu_{{\rm max},\odot})(T_{\rm eff}/{\rm T}_{{\rm eff},\odot})^{1/2},  
\end{eqnarray}
where
$\nu_{{\rm max},\odot} = 3050$~$\mu$Hz, $\Delta\nu_{\odot} = 134.92$~$\mu$Hz,
${\rm T}_{{\rm eff},\odot} = 5777$~K, and ${\rm g}_{\odot} = 10^{4.44}$~cm~s$^{-2}$.
Such recalculated $R_{\rm seis}$, $M_{\rm seis}$, and $\log g_{\rm seis}$ for each star
are given in Table 1, and the variations relative to Mosser et al.'s (2012) original values
are shown in Fig.~10. We can see from this figure that the differences in $\log g_{\rm seis}$
are very marginal (0.01--0.02~dex in most cases) and practically negligible,
because $g_{\rm seis}$ is insensitive to $T_{\rm eff}$ (i.e., $\propto T_{\rm eff}^{1/2}$).

\setcounter{figure}{9}
\begin{figure*}
\begin{minipage}{140mm}
\begin{center}
\includegraphics[width=14.0cm]{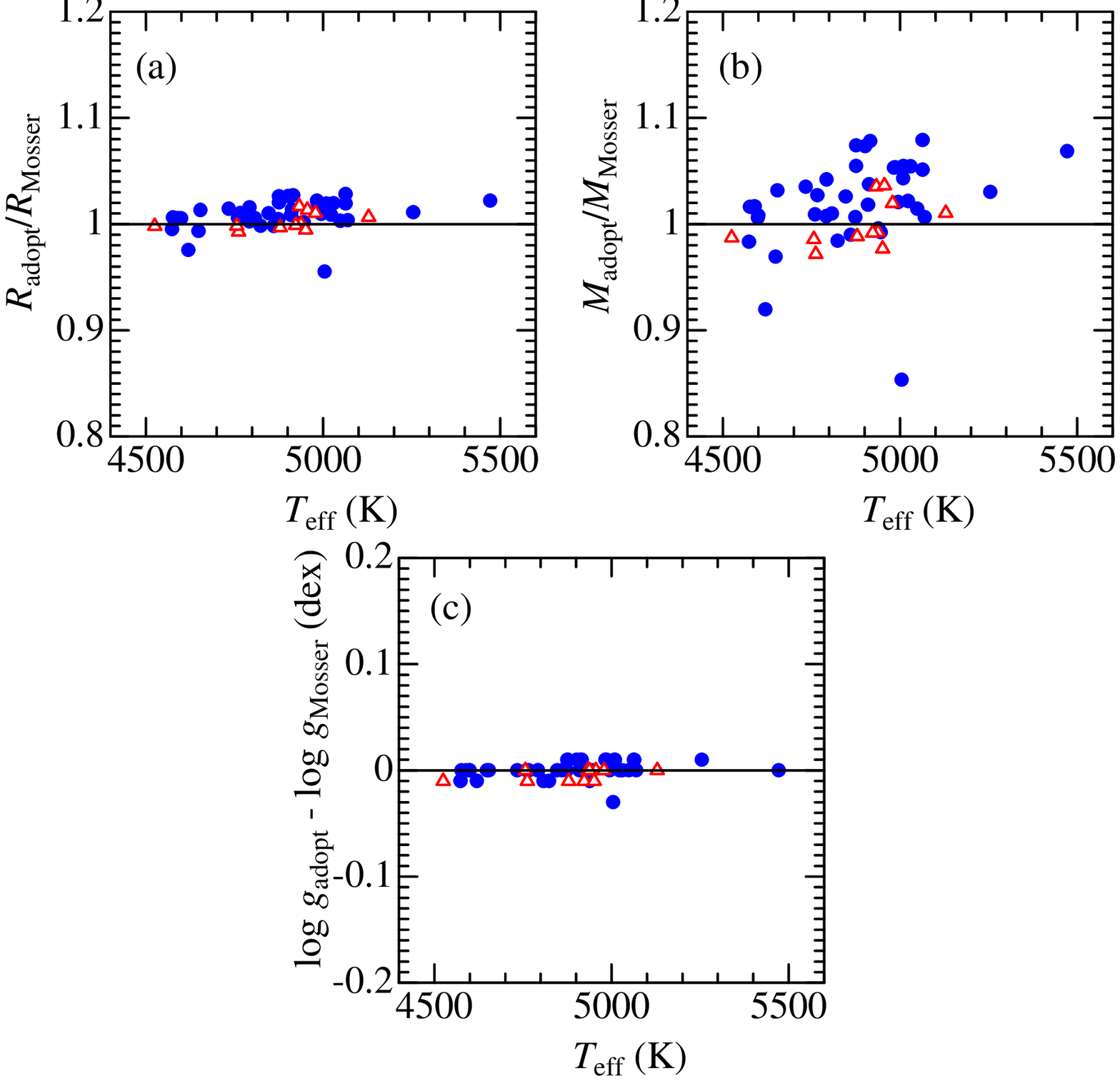}
\caption{Variations between the finally adopted seismic $R$, $M$, and $\log g$ 
[obtained by applying our spectroscopic $T_{\rm eff}$ to Eqs. (1)--(3)] 
and Mosser et al.'s (2012) original values [based on $T_{\rm eff}$(KIC)],
plotted against $T_{\rm eff}$. The meanings of the symbols are the same as in 
Fig. 7.}
\label{fig10}
\end{center}
\end{minipage}
\end{figure*}

\setcounter{figure}{10}
\begin{figure}
\begin{minipage}{80mm}
\begin{center}
\includegraphics[width=7.0cm]{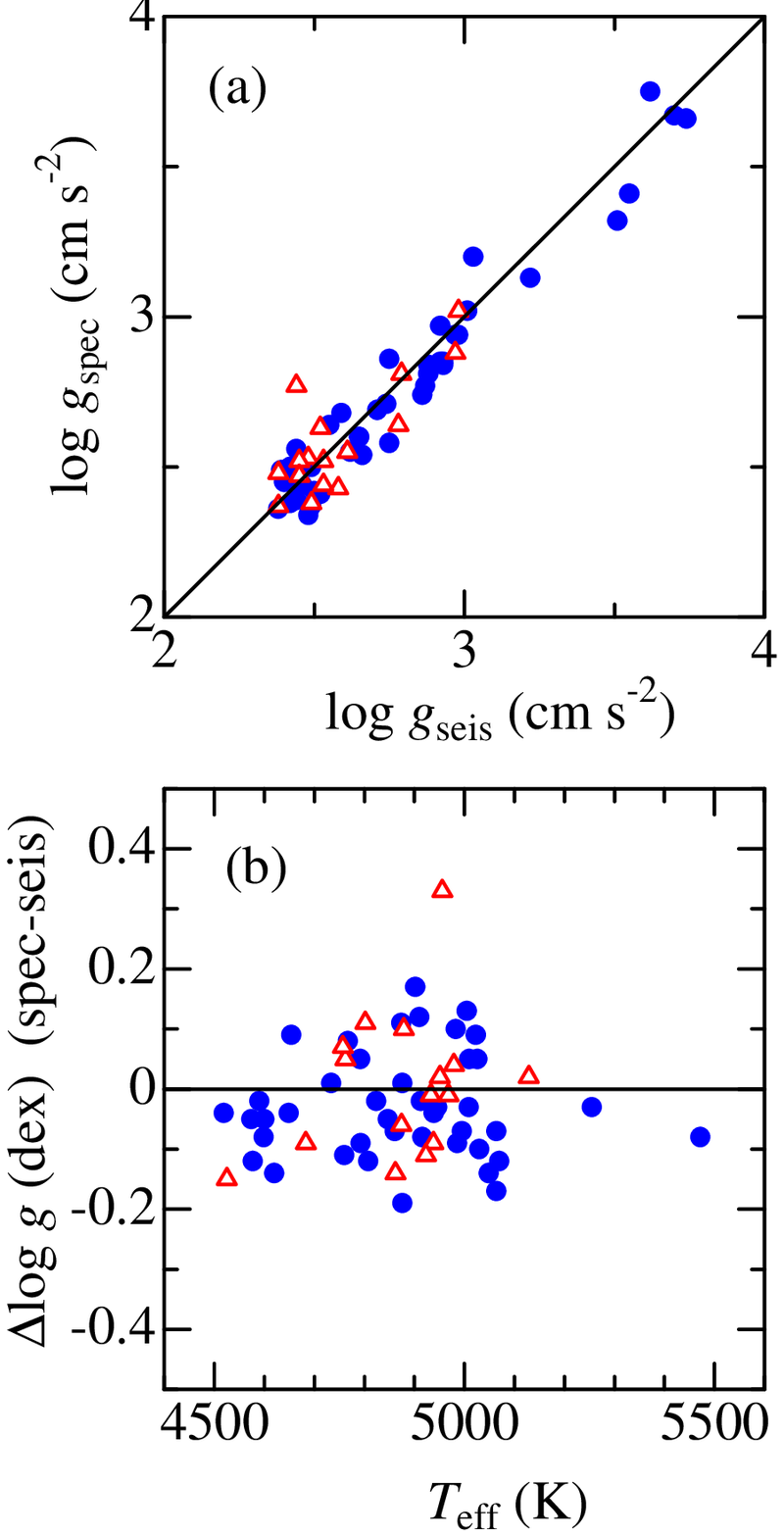}
\caption{(a) Comparison of our spectroscopic $\log g_{\rm spec}$ with 
the seismic $\log g_{\rm seis}$. (b) Plot of the $\log g$ difference 
[$\Delta \log g (\equiv \log g_{\rm spec} - \log g_{\rm seis})$] 
against $T_{\rm eff}$.  The meanings of the symbols are the same 
as in Fig. 7.
}
\label{fig11}
\end{center}
\end{minipage}
\end{figure}

The comparison between $\log g_{\rm spec}$ and $\log g_{\rm seis}$ is displayed
in Fig.~11a, and the difference $\Delta\log g$ ($\equiv \log g_{\rm spec} - \log g_{\rm seis}$) 
is plotted against $T_{\rm eff}$ in Fig.~11b. We can see from these figures that 
both are in satisfactory agreement with each other ($r = 0.96$; cf. Table 3). 
Actually, the mean difference 
averaged over all 58 stars is $\langle \Delta \log g\rangle$ is $-0.02$~dex 
with the standard deviation of $\sigma$ = 0.10~dex.\footnote{If the results from two
different data sets are to be separately treated, we obtain  
$\langle \Delta \log g\rangle = -0.03$ ($\sigma$ = 0.09) for 42 stars based on 
our Subaru data, and $\langle \Delta \log g\rangle = +0.01$ ($\sigma$ = 0.12) 
for 16 stars based on Thygesen et al.'s (2012) spectra.}
Therefore, we may conclude that our spectroscopic $\log g$ determination is 
sufficiently reliable, in the sense that it can be used for evaluating absolute 
$\log g$ values of red giants to a precision of $\simeq 0.1$~dex.   
 
It should be remarked, however, that this conclusion applies to the 
spectroscopic $\log g$ obtained by using the TGVIT program as done in Paper I.  
Given that the $\log g$ values spectroscopically derived 
in Paper I were systematically smaller than those of several other previous studies
as already mentioned in Sect. 1, we may state based on the present conclusion 
that those ``high-scale'' $\log g$ results published by previous investigators 
are likely to be inadequately overestimated despite that they were derived 
in a similar spectroscopic method using Fe~{\sc i} and Fe~{\sc ii} lines. 
Besides, although Thygesen et al. (2012) arrived at a similar result 
that the spectroscopic and seismic $\log g$ values are almost consistent
with each other in the average sense, the mean difference was 
$\langle \Delta \log g\rangle = -0.05$ ($\sigma$ = 0.30) for 57 stars, 
which means that the dispersion of deviation in their results is 3 times 
as large as ours (cf. Fig. 3 in their paper).\footnote{
In contrast to Thygesen et al.'s (2012) implication, we consider that LTE is 
practically valid in the Fe~{\sc i}/Fe~{\sc ii} ionization equilibrium 
as far as red giant stars under study are concerned, since the
deviation $\Delta \log g$ does not show any $T_{\rm eff}$-dependence
(cf. Fig.~11b).}
Thus, since the effectiveness of spectroscopic parameter determinations 
appears to be rather case-dependent, one should keep more attention to 
practical details involved with his method (e.g., which quality/strength 
and how many lines are to be used, atomic parameters, criteria level of 
required conditions). 

\subsection{Mass problem in red clump giants}

In Figs.~12a and 12b are compared $M_{\rm trk}$ (mass estimated from evolutionary 
tracks) and $R_{\rm ph}$ (radius evaluated from $L$ and $T_{\rm eff}$) derived 
for 9 stars with known parallax data (cf. Table 2) to the corresponding seismic 
$M_{\rm seis}$ and $R_{\rm seis}$, respectively.
We see from Fig. 12b that $R_{\rm ph}$ and $R_{\rm seis}$ are almost in 
agreement except for KIC~09705687, for which the result is unreliable 
because its parallax contains a large error. 
In contrast, Fig. 12a reveals that $M_{\rm trk}$ is considerably larger
than $M_{\rm seis}$ roughly by $\sim 50$\% on the average 
(i.e., $\sim$~1--2~$M_{\rm seis}$ corresponds to $\sim$~1.5--3~$M_{\rm trk}$). 
However, since $M_{\rm trk} \simeq M_{\rm seis}$ holds for KIC~10323222 
(normal red giant; RG), appreciable discrepancies are likely to be seen
only in red clump giants (RC1/RC2). 
These facts suggest that our suspicion mentioned in Sect. 1 resulting 
from the comparison of Figs. 2b and 2e was surely correct. That is,
we must now admit that the $M_{\rm trk}$ values for red clump giants
(roughly corresponding to the intersection of the trifid structure 
in Fig. 2e) derived in Paper I were erroneously overestimated. 

\setcounter{figure}{11}
\begin{figure}
\begin{minipage}{80mm}
\begin{center}
\includegraphics[width=7.0cm]{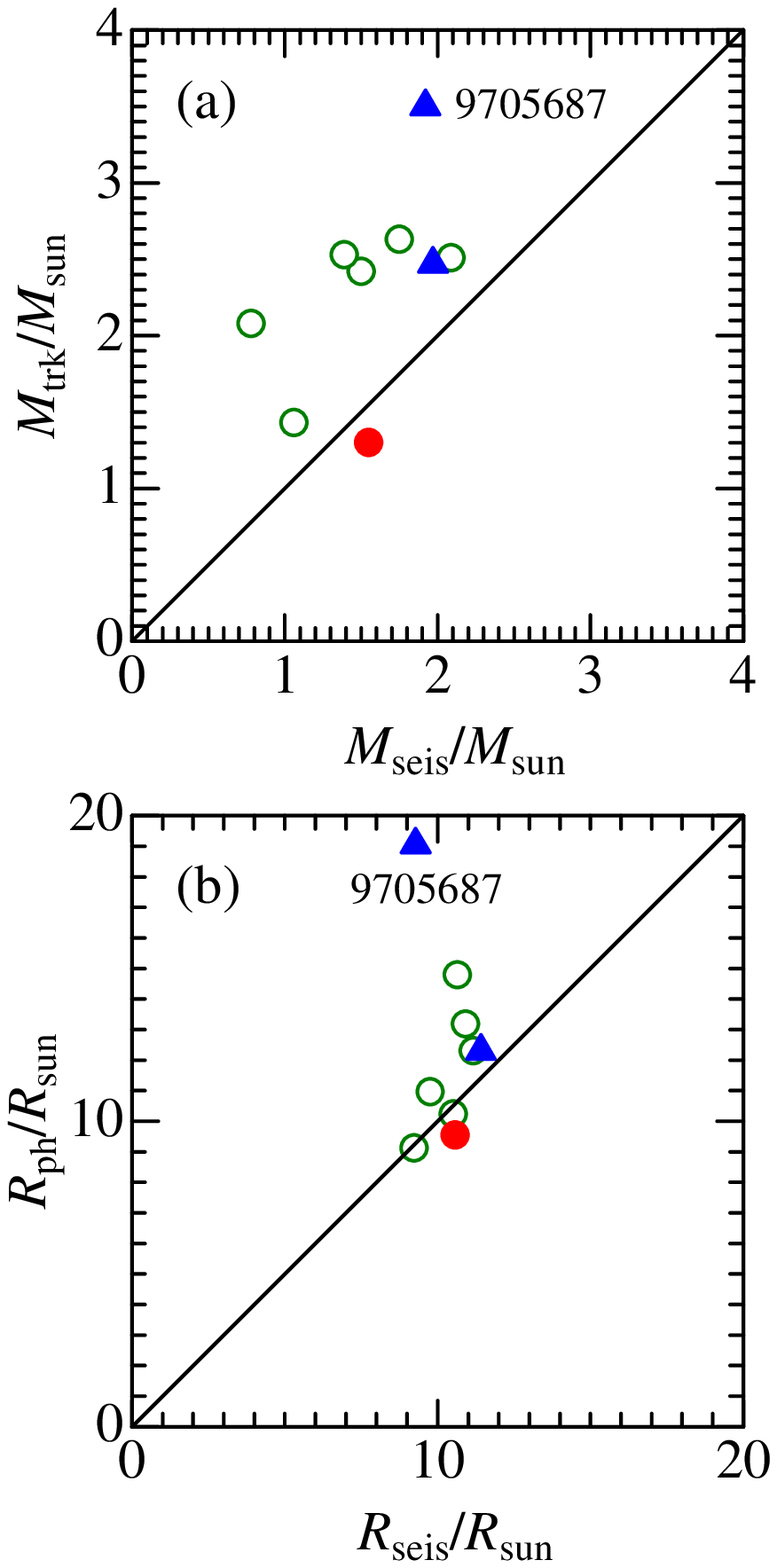}
\caption{Comparison of seismic mass or radius with those evaluated 
in the conventional approach for 9 stars, for which parallax data are
available (see Table 2).  Upper panel (a): $M_{\rm trk}$ (mass estimated 
from evolutionary tracks) vs. $M_{\rm seis}$ (seismic mass). 
Lower panel (b): $R_{\rm ph}$ (photometric radius derived from $L$ 
and $T_{\rm eff}$) vs. $R_{\rm seis}$ (seismic radius).
See the caption of Fig. 2 for the meanings of the symbols.}
\label{fig12}
\end{center}
\end{minipage}
\end{figure}

Presumably, the cause for this disagreement is related to the process 
of mass determination adopted in Paper I, where Lejeune \& Schaerer's (2001)
grids of stellar evolutionary tracks (computed for 11 $M$ values of 0.8, 0.9, 
1.0, 1.25, 1.5, 1.7, 2.0, 2.5, 3.0, 4.0, and 5.0~M$_{\odot}$ for each of the 
six metallicities $z$ = 0.001, 0.004, 0.008, 0.02, 0.04, and 0.10) were used.
That is, for a set of evolutionary tracks corresponding to each mass $M_{i}$, 
the minimum luminosity difference ($\delta_{i} \equiv \log L_{i} - \log L^{*}$, 
where $\log L^{*}$ is the given luminosity of a star) was computed\footnote{
If several different $\delta$ values exist at a fixed $T_{\rm eff}$ 
(i.e., for the case of a looped track), we adopt the one corresponding to the smallest
$|\delta|$ among those.} at the interpolated point(s) on the relevant track 
for $T_{\rm eff} = T_{\rm eff}^{*}$. Then, the best $M$ value (for the metallicity 
of the relevant tracks) accomplishing the minimum $|\delta|$ was estimated 
by interpolation based on the set of \{$\delta_{i}$, $i$ = 1,11\}. 
This whole process was repeated for each set of tracks corresponding to 
different metallicities, and the final $M$ value corresponding to the observed 
stellar [Fe/H] was determined again by interpolation.

In this way of $M$ determination for a star at the red clump luminosity  
of $\log (L/{\rm L_{\odot}}) \sim$~1.5--2, the result tends to be the one 
corresponding to the normal red giant evolution (even if it is really 
a red clump star), because a solution of $M$ (for any [Fe/H]) accomplishing 
the given stellar luminosity is always found for such red giant tracks 
running nearly horizontally and discriminated by ($M$, [Fe/H])  (cf.  Fig. 2d). 
Thus, $M \sim$~2--3~M$_{\odot}$ would naturally obtained for those stars 
in the red clump zone (cf. Fig. 2e).
However, actual masses of those red-clump stars in the He-burning stage
to be found at $\log (L/{\rm L_{\odot}}) \sim$~1.5--2, which have returned 
from the RGB tip, are much smaller to be $\sim$~1--2~M$_{\odot}$ (see, e.g., 
Fig. 10 in Kallinger et al. 2010b). This must be the reason for the discrepancy 
between $M_{\rm trk}$ ($\sim$~1.5--3~M$_{\odot}$) and $M_{\rm seis}$
($\sim$~1--2~M$_{\odot}$) we found for red clump giants in Fig. 12a.

Interestingly, the consequence that the $M$ values derived in Paper I are likely 
to be significantly overestimated for red clump giants (constituting 
a large fraction of 322 targets) provides a reasonable solution for 
the inexplicable features noted in that paper. \\
-- First, the surface gravity ($\log g_{TLM}$) directly derived from $T_{\rm eff}$,
$L$, and $M$ was found to be systematically larger by $\sim$~0.1--0.2~dex 
than the spectroscopic gravity ($\log g_{\rm spec}$) just around $\log g \sim$~2--3 
corresponding the red clump region (cf. Fig. 1g therein). This discrepancy can be 
naturally explained by an overestimation of $M$ by $\sim 50$\%.\\
-- Second, it was remarked in Paper I that the $age$ vs. [Fe/H] relation 
derived for 322 giants and that for 160 FGK dwarfs obtained by Takeda (2007) 
do not connect smoothly but show an appreciable discontinuity especially around 
$age \sim 10^{9}$~yr (cf. Fig.~13, which is equivalent to Fig. 14 in Paper I) 
That is, while the large scatter of [Fe/H]$_{\rm dwarfs}$ seen in 
old stars ($age \sim 10^{10}$~yr) tends to converge toward medium-aged 
stars ($age \sim 10^{9}$~yr), a large spread reappears in [Fe/H]$_{\rm giants}$ 
at $age \sim 10^{9}$~yr which again shrinks toward young stars ($age \sim 10^{8}$~yr).
However, we now know that many red clump giants should have masses around 
$\sim 1.5$~M$_{\odot}$ (corresponding to $\log age \sim 9.3$; cf. Fig. 3c in Paper I) 
instead of the previous values around $\sim 2.5$~M$_{\odot}$ ($\log age \sim 8.8$).
It is gratifying to see that the increase in the age of clump giants by 
$\Delta\log age \sim 0.5$~dex corresponding to this revision satisfactorily removes 
the discrepancy between two $age$ vs. [Fe/H] distributions (cf. the arrow in Fig.~13).

\setcounter{figure}{12}
\begin{figure}
\begin{minipage}{80mm}
\begin{center}
\includegraphics[width=8.0cm]{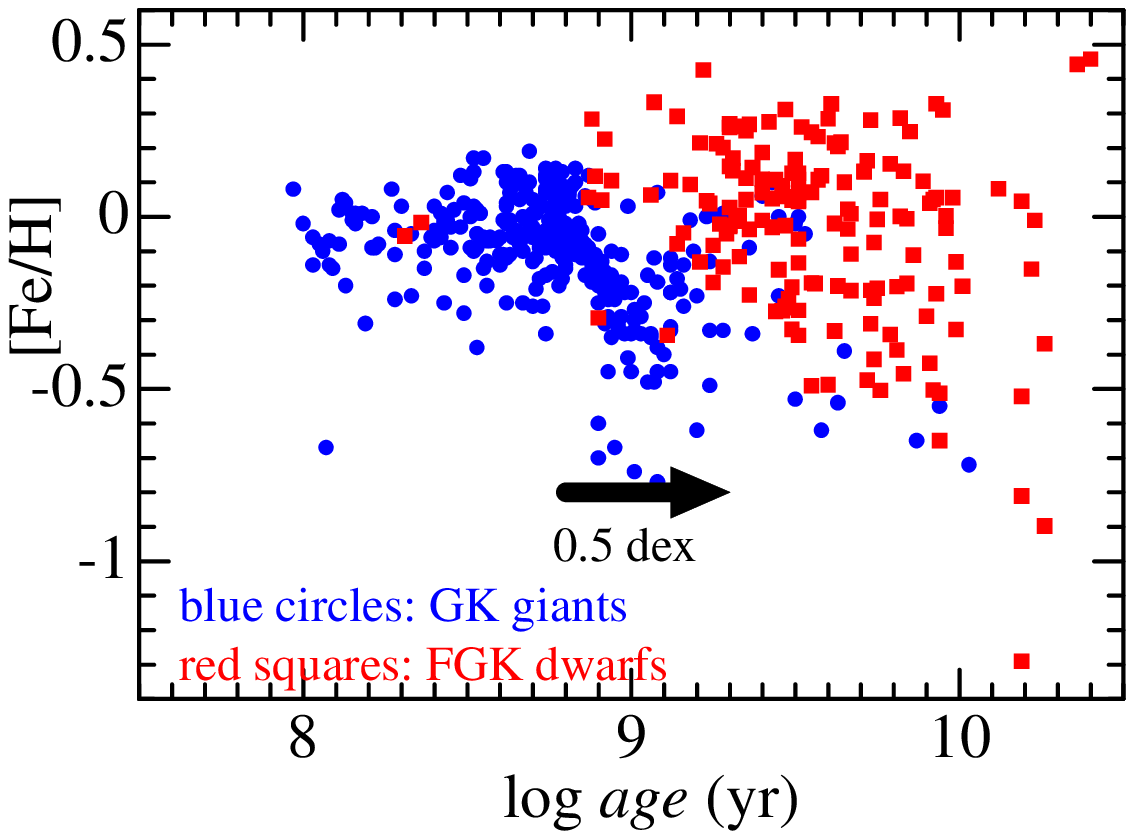}
\caption{Metallicity ([Fe/H]) vs. $age$ relation for
322 G-K giants (circles) based on the results of Paper I
and for 160 FGK dwarfs based on the results of Takeda (2007).
The arrow indicates the increase of age to be applied to red clump giants 
(from $\log age \sim 8.8$ to $\log age \sim 9.3$ by $\sim 0.5$~dex)
resulting from the downward revision of their masses
(cf. the last paragraph in Sect. 4.2) 
Note that this figure is essentially the same as Fig. 14 in Paper I.}
\label{fig13}
\end{center}
\end{minipage}
\end{figure}

\subsection{Classification of red giants}

Finally, let us discuss the question inspired by the similarity of Figs. 2b and 2e:
``Is it possible to distinguish the different evolutionary status of red giants 
(RG/RC1/RC2) only by way of the conventional approach as adopted in Paper I?''
We now know that the spectroscopic $\log g$ values are sufficiently reliable, 
while the $M$ values estimated from theoretical tracks as done in Paper I are 
erroneously too large typically by $\sim 50$\% for clump giants (RC1/RC2) 
though they are still reasonably usable for normal H-burning red giants (RG).
Keeping this information in mind, we can roughly figure out the classification 
of many red giants plotted in Fig. 2e (gray symbols).
For example, a bunch of stars located in the region of lower $M$ but higher $\log g$  
($1 \la M_{\rm trk}/{\rm M}_{\odot} \la 2$ and $2.5 \la \log g_{\rm spec} \la 3.5$) 
presumably belong the RG class. Most of the stars densely populated
in the heart of Fig. 2e ($2 \la M_{\rm trk}/{\rm M}_{\odot} \la 3$ and 
$2 \la \log g_{\rm spec} \la 3$) are likely to be in the red clump (RC1/RC2)
category, despite that their $M_{\rm trk}$ values are systematically overestimated
compared to the true values. But unfortunately, reliable classification for each star 
in this crowded clump region is very difficult, especially regarding the distinction 
of RC1 and RC2 where information of absolute $M$ values is required, 
as $M_{\rm seis} = 1.8 M_{\odot}$ was adopted by Mosser et al. (2012) 
for the distinction limit between these two. 
Since this critical mass value almost coincides with the intersection point of 
``trifid'' structure in Fig. 2b, the corresponding value in Fig. 2e might be 
$M_{\rm trk} \sim 2.5$~M$_{\odot}$ where the intersection in the similar distribution is seen.  
However, the real situation should not be so simple as such a naive analogy, since 
the required correction in $M_{\rm trk}$ must be different between RG and RC1/RC2,
which makes us rather cautious in placing too much weight on the similarity of 
Figs. 2b and 2e.   
Thus, we have to conclude that the approach of Paper I (at least in the present 
level) can not be so effective as the asteroseismic technique in distinguishing 
the evolutionary status of red giants from the viewpoint of accuracy.\footnote{
In connection with this conclusion, we first expected that $v_{\rm e}\sin i$
may be used to discriminate between RG and RC1/RC2, since their different evolution 
histories might have resulted in different degree of angular momentum loss. 
While we notice in Figs. 9a and 9b that $v_{\rm e}\sin i$(RG) tend to be  
smaller than $v_{\rm e}\sin i$(RC1/RC2), we feel it still premature to take this 
trend as very meaningful, since RG-class stars in our sample are biased to 
smaller masses ($M_{\rm seis} \sim$~1--2~M$_{\odot}$). That is, this apparent 
trend may be simply due to the $M$-dependence of $v_{\rm e}\sin i$.
In order to check whether or not this tendency really reflects the evolution-related 
effect, the sample would have to be further increased so that stars of all three 
classes may be available over a sufficiently wide range of $M$.} 
   
Nevertheless, we consider that this conventional procedure (spectroscopic method 
coupled with the help of stellar evolutionary tracks) may possibly be improved to 
a practicable level, though this assumes that reliably accurate parallax data 
are available as a prerequisite.
For example, we would be able to derive two kinds of $M_{\rm trk}$ for a star 
locating in the clump region on the HR diagram while {\it assuming} its status
(i.e., either H-burning RG or He-burning RC) in advance, where much more refined
evolutionary tracks with finer mass steps (e.g., 0.1~M$_{\odot}$) should be used
such as shown in Fig. 10 of Kallinger et al. (2010b).
Which of these two solutions of $M_{\rm trk}$(RG) and $M_{\rm trk}$(RC) 
represents the truth may be judged from the $\log g_{\rm spec}$ vs. $R_{\rm ph}$
relation where RG and RC tend to dwell rather separately (cf. Fig. 2f in 
comparison with Fig. 2c), because we may hope that these two quantities can be 
well established (given that accurately known distance allows a reliable evaluation 
of $L$). Or alternatively, this decision could be made by checking
the consistency between $\log g_{TLM}$ and $\log g_{\rm spec}$ (cf. Sect. 4.2).
Then, if $M_{\rm trk}$(RC) turned out to be the correct solution, the distinction
between RC1 and RC2 is straightforward from its comparison with the critical mass 
($\sim$~1.8~M$_{\odot}$).
 
In any event, availability of reliable parallax data is essential for such 
a method of approach to work successfully, since precisely evaluated $L$ 
(along with spectroscopically determined $T_{\rm eff}$ and [Fe/H])
is necessary for correctly setting the position of a star on the $\log L$ vs. 
$\log T_{\rm eff}$ diagram, which should be compared with theoretical tracks
to determine $M$. It is unfortunate that parallaxes are currently known (even so, 
the accuracy is not yet sufficient) for only a limited number of Mosser et al.'s (2012) 
{\it Kepler} giants with asteroseimologically established classification 
and seismic parameters (e.g., only 9 out of 55 targets in the present case).
We expect, however, that this situation will soon be significantly improved
by the high-precision parallax data to be released by Gaia.
After very precise distance data have been established for all these 
red giants with known seismic properties, we would like to revisit this problem  
by carrying out a renewed analysis similar to this study (hopefully based 
on more extended samples), in order to see whether such an approach of 
diagnosing the nature of red giants we propose efficiently works out.    

\section{Summary and conclusion}

Recent progress in asteroseismology based on very high-precision photometry 
from satellites such as {\it Kepler} has enabled to successfully sort out the
various complex evolutionary stages of many red giants [either normal H-burning 
giants (RG) ascending the RGB or He-burning giants (RC1/RC2) residing in the 
1st or 2nd clump after returning from the RGB tip)] as well as to precisely 
determine their mass and radius.

When the correlations of such seismic parameters ($M_{\rm seis}$, $R_{\rm seis}$, 
and $\log g_{\rm seis}$) established by Mosser et al. (2012) for $\sim 200$ red giants 
observed by {\it Kepler} were compared with those previously determined in Paper I 
for a large number of field G--K giants based on the conventional approach 
(i.e., spectroscopic $T_{\rm eff}$, $\log g$, $v_{\rm t}$, and [Fe/H] 
based on the equivalent widths of Fe~{\sc i} and Fe~{\sc ii} lines, while 
$M$ and $R$ evaluated from $L$ with the help of stellar evolutionary tracks), 
we noticed an interesting similarity in the $\log g$ vs. $M$ diagram, 
where the seismic parameters of different evolutionary classes show 
characteristic distributions, which implied that even such an ordinary approach 
as adopted in Paper I might possibly be used for distinguishing the complex 
evolutionary stages of red giants.  

As a first step toward this possibility, we decided to examine the precision of 
conventionally derived $\log g$ and $M$ in reference to the seismic values.
For this purpose, we applied the same parameter determination method 
as adopted Paper I to selected 55 red giants in the $Kepler$ field, for which 
seismic parameters and evolutionary classes are well defined by Mosser et al. (2012).
 
The spectroscopic parameters were derived by using the TGVIT program 
(Takeda et al. 2005) with the equivalent widths of Fe~{\sc i} and Fe~{\sc ii} lines 
measured mainly on the high-dispersion spectra obtained by Subaru/HDS (42 stars) 
while partly on the spectra published by Thygesen et al. (2012) (16 stars). 
We also derived $v_{\rm e}\sin i$ (projected rotational velocity) from 
the line-broadening width by a spectrum-fitting analysis in the 6080--6089~$\rm\AA$ 
region as in Paper I.  

It was confirmed that our spectroscopic gravity ($\log g_{\rm spec}$) and the seismic 
gravity ($\log g_{\rm spec}$) are in satisfactory agreement with each other (to within
$\simeq 0.1$~dex) without any systematic difference, which ensures the reliability
of our spectroscopic parameters based on Fe~{\sc i} and Fe~{\sc ii} lines.

For 9 stars for which parallax data are available among these spectroscopic targets,
the stellar parameters relevant to the HR diagram ($L$, $M_{\rm trk}$, and 
$R_{\rm ph}$) were also determined. While we found a reasonable consistency
between $R_{\rm ph}$ and $R_{\rm seis}$ (except for 1 star with an unreliable parallax),
the masses of He-burning red clump giants derived from evolutionary tracks turned out 
to be $M_{\rm trk}({\rm RC1/RC2}) \sim$~2--3~M$_{\odot}$, which are markedly larger by 
$\sim$~50\% than the seismic results of $M_{\rm seis}({\rm RC1/RC2}) \sim$~1--2~M$_{\odot}$, 
though such discrepancy is not apparent for normal H-burning giants 
[$M_{\rm trk}({\rm RG}) \simeq M_{\rm seis}({\rm RG})$]. 
This disagreement is presumably attributed to the difficulty of mass determination 
from the evolutionary tracks where RG paths (higher $M$) and RC paths (lower $M$) 
are intricately overlapping at the same clump region of the HR diagram; i.e., 
simply running RG tracks tend to be preferentially used even for RC stars. 

This consequence naturally implies that many of the $M_{\rm trk}$ values of 
322 red giants (most of them are likely to be red clump stars) estimated in Paper I 
were erroneously overestimated by an order of $\sim$~50\%, about which serious caution 
should be taken, though other results (e.g., $R_{\rm ph}$ values or spectroscopic 
parameters) derived therein do not have any problem. 
The fact that an upward revision by $\sim$~50\% for the mass values in Paper I 
removes the puzzling discrepancy found in Paper I (i.e., systematic difference 
between $\log g_{\rm spec}$ and $\log g_{TLM}$, a mismatch between dwarfs and giants 
in the distribution of [Fe/H] vs. $age$ diagram) substantiates this conclusion.

We must conclude that the traditional approach of Paper I is not so effective 
as the asteroseismic technique in distinguishing the evolutionary status of red giants.
However, it may be still useful (at least for guessing the evolutionary class)
if parallaxes are reliably determined (for accurate estimation of $L$) 
and up-to-date theoretical tracks (covering the whole RG--RC stage with a 
sufficiently small $M$ step) are used, since we may hope that $T_{\rm eff}$, 
$\log g$, and [Fe/H] can be reliably determined spectroscopically.
Above all, the availability of precise distances is essential.
When very accurate parallax data have become available in the near future 
for the reference giants with known seismic properties, it would be worth 
carrying out a renewed analysis similar to this study, in order to check  
whether such a conventional approach for diagnosing the nature of red giants
really works out.   

\section*{Acknowledgments}

This research has been carried our by using the SIMBAD database,
operated by CDS, Strasbourg, France.

\appendix
\section{Recent high-resolution grid of stellar evolutionary tracks}

We discussed in Sect. 4.3 the prospect of reliably determining the mass 
of red giants, where we suggested to derive two kinds of stellar masses 
by assuming its evolutionary status in advance (i.e., either RG or RC) 
and then adopt the more reasonable alternative. 
But this assumes the application of evolutionary tracks with sufficiently 
fine parameter grid as prerequisite, since those used in Paper~I 
($\Delta M/M \sim$~20--30\%, $\Delta \log z \sim$~0.3--0.4~dex; cf. Sect. 4.2)
were evidently not satisfactory in this respect. How is the current
situation regarding this point?

Fortunately, thanks to the remarkable progress in this field, such 
theoretical tracks of sufficiently high-resolution grid have already 
been published by several groups. For example, the BaSTI database\footnote{
Available from {\tt http://albione.oa-teramo.inaf.it/}
}
(Pietrinferni et al. 2004) provides tracks with grids of  
$\Delta M/M \sim$~5--10\% (for $\sim$~1--4~M$_{\odot}$ stars) and 
$\Delta \log z \sim$~0.1--0.2~dex (for the metallicity range of disk population),
which were used by Kallinger et al. (2010b) and proved to be useful for
discussing the parameters of red giants. 
More noteworthy is the recent contribution of the Padova--Trieste group, 
who published extensive data of evolutionary tracks and isochrones\footnote{
Available from {\tt http://stev.oapd.inaf.it/parsec\_v1.0/}
or {\tt http://people.sissa.it/\~{ }sbressan/parsec.html}
}
computed based on the PARSEC code (Bressan et al. 2012, 2013) with generally very 
fine grids of $\Delta M/M \sim$~2--5\%  (0.05~M$_{\odot}$ step for 1--2.3~M$_{\odot}$ 
and 0.1~M$_{\odot}$ step for 2.3--5~M$_{\odot}$) and $\Delta \log z \sim$~0.1~dex. 
Even results of further finer mass step (0.025~M$_{\odot}$) are provided around 
1.7--2~M$_{\odot}$, which makes this database especially suitable for revealing 
the complex behavior of tracks in the red-clump region. 

As a demonstration, these PARSEC tracks of near-solar metallicity ($z=0.02$; 
Z0.02Y0.284.tar.gz) in the mass range 1--4~M$_{\odot}$ (with $\Delta M = 0.05$~M$_{\odot}$ 
in 1--2.3~M$_{\odot}$, while $\Delta M = 0.1$~M$_{\odot}$ in 2.3--4~M$_{\odot}$) 
around the red-clump region are depicted in Fig. A1a/a$'$ (before He ignition) 
and Fig. A1b/b$'$ (after He ignition).
For comparison, the tracks of Lejeune \& Schaerer (2001), which were used for 
evaluating $M_{\rm trk}$ in Paper~I as well as in Sect. 3.3, are shown in Fig. A1c 
and Fig. A1d, each corresponding to the phase of H-buring and He-burning, respectively. 

We can immediately recognize from these figures that considerable improvements 
in accuracies of mass determination are expected by using these recent PARSEC tracks,
in comparison to the previous case of Paper I where coarse tracks were used.
The distinct merit is that we do not have to worry any more about difficulties and 
uncertainties involved in interpolating (or extrapolating) the grid of tracks. 
Of course, since actual tracks are not simple and complexities (e.g., 
looped or bumped features) more or less exist especially at the post He-ignition 
phase (Fig. A1b/b$'$), some uncertainties are naturally inevitable. Yet, we presume 
that $M$ would be determinable to a precision of $\la 10$\% (given that 
whether pre- or post-He ignition is presupposed) by using these evolutionary 
tracks in the near future, when $L$ has been reliably established based on the 
very accurate parallax data to be released by Gaia, while $T_{\rm eff}$ and $z$ are 
expected to be settled spectroscopically with sufficient accuracies ($< 0.01$~dex in 
$\log T_{\rm eff}$ and $< 0.1$~dex in $\log z$). 

\setcounter{figure}{0}
\begin{figure*}
\begin{minipage}{140mm}
\begin{center}
\includegraphics[width=14.0cm]{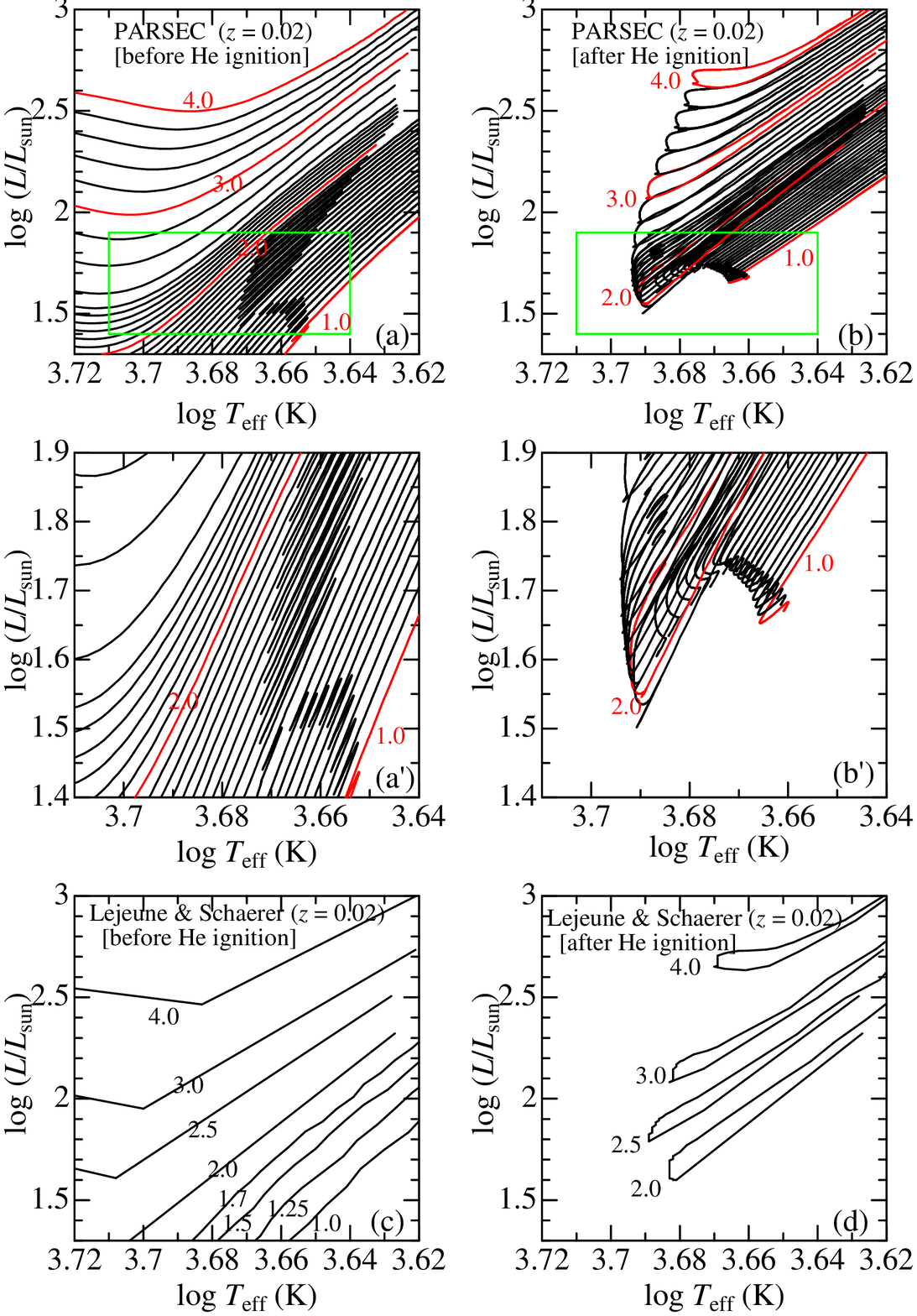}
\caption{
Top two panels: Theoretical evolutionary tracks for 
1--4~M$_{\odot}$ stars of near-solar metallicity ($z = 0.02$) around the red-clump 
region with high-resolution mass steps (0.05~M$_{\odot}$ for 1--2~M$_{\odot}$ stars, 
0.2~M$_{\odot}$ for 2--4~M$_{\odot}$ stars), which were calculated by the Padova--Trieste 
group with the PARSEC code (Bressan et al. 2012, 2013). In the left panel (a) are shown 
the tracks corresponding to the shell H-burning phase before He ignition, while 
the right panel (b) presents those of the core He-burning phase after He ignition.
The tracks for 1, 2, 3, and 4 M$_{\odot}$ are colored in red for convenience. 
Middle two panels: The left (a$'$) and right (b$'$) panels show magnification of 
the red-clump area indicated by green rectangles in the upper panels (a) and (b), respectively. 
Bottom two panels: 
Lejeune \& Schaerer's (2001) $z = 0.02$ tracks with rather coarse mass grid (which 
were used for mass determination in Paper~I as well as in Sect. 3.3 in this paper).
The left panel (c) corresponds to the pre-He ignition phase, while the right panel (d) 
to the post-He ignition.
}
\label{figA1}
\end{center}
\end{minipage}
\end{figure*}

\end{document}